\documentclass[12pt]{article}
\usepackage{epsf}
\usepackage{amsmath}
\usepackage{amsfonts}
\usepackage{amssymb}
\usepackage{graphicx}
\usepackage{color}
\usepackage{psfrag}
\usepackage{cite}

\usepackage{ifpdf}

\newcommand{\bmat}{\left(\begin{array}}
\newcommand{\emat}{\end{array}\right)}

\def\yzero{\smash{\hbox{$y\kern-4pt\raise1pt\hbox{${}^\circ$}$}}}

\def\a{\alpha}
\def\b{\beta}

\def\beq{\begin{equation}}
\def\eeq{\end{equation}}
\def\beqa{\begin{eqnarray}}
\def\eeqa{\end{eqnarray}}

\def\-{\hphantom{-}}
\def\ov{\overline}
\def\s2{\frac{1}{\sqrt2}}

\def\beq{\begin{equation}}
\def\eeq{\end{equation}}
\def\beqa{\begin{eqnarray}}
\def\eeqa{\end{eqnarray}}

\def\IF{\relax{\rm I\kern-.18em F}}
\def\II{\relax{\rm I\kern-.18em I}}

\def\Dsl{\,\raise.15ex\hbox{/}\mkern-13.5mu D} 

\def\IS{{\bf S}}
\def\IR{{\bf R}}
\def\IZ{{\bf Z}}
\def\IX{{\bf X}}
\def\IB{{\bf B}}
\def\IT{{\bf T}}

\catcode`\@=11   
\newdimen\@rotdimen
\newbox\@rotbox  

\def\@vspec#1{\special{ps:#1}}
\def\@rotstart#1{\@vspec{gsave currentpoint currentpoint translate
   #1 neg exch neg exch translate}}
\def\@rotfinish{\@vspec{currentpoint grestore moveto}}
\def\@rotr#1{\@rotdimen=\ht#1\advance\@rotdimen by\dp#1%
   \hbox to\@rotdimen{\hskip\ht#1\vbox to\wd#1{\@rotstart{90 rotate}%
   \box#1\vss}\hss}\@rotfinish}
\def\@rotl#1{\@rotdimen=\ht#1\advance\@rotdimen by\dp#1%
   \hbox to\@rotdimen{\vbox to\wd#1{\vskip\wd#1\@rotstart{270 rotate}%
   \box#1\vss}\hss}\@rotfinish}%
\def\@rotu#1{\@rotdimen=\ht#1\advance\@rotdimen by\dp#1%
   \hbox to\wd#1{\hskip\wd#1\vbox to\@rotdimen{\vskip\@rotdimen
   \@rotstart{-1 dup scale}\box#1\vss}\hss}\@rotfinish}%
\def\@rotf#1{\hbox to\wd#1{\hskip\wd#1\@rotstart{-1 1 scale}%
   \box#1\hss}\@rotfinish}%
\def\rotate{\@ifnextchar[{\@rotate}{\@rotate[l]}}
\def\@rotate[#1]#2{\setbox\@rotbox=\hbox{#2}\@nameuse{@rot#1}\@rotbox}

\catcode`\@=12

\topmargin
-1.5cm
\textwidth
15.5cm
\textheight
23.5cm
\oddsidemargin
0.7cm
\evensidemargin
1.2cm

\begin{document}

\makeatletter
\@addtoreset{equation}{section}
\makeatother
\renewcommand{\theequation}{\thesection.\arabic{equation}}
\pagestyle{empty}
\rightline{ IFT-UAM/CSIC-12-108}
\vspace{0.1cm}
\begin{center}
\LARGE{\bf $\IZ_p$ charged branes \\in flux compactifications \\[12mm]}
\large{M. Berasaluce-Gonz\'alez$^{1,2}$, P. G. C\'amara$^{3,4}$, \\F. Marchesano$^{2}$, A. M. Uranga$^2$\\[3mm]}
\footnotesize{${}^{1}$ Departamento de F\'{\i}sica Te\'orica,\\[-0.3em] 
Universidad Aut\'onoma de Madrid, 28049 Madrid\\
${}^2$ Instituto de F\'{\i}sica Te\'orica IFT-UAM/CSIC,\\[-0.3em] 
C/ Nicol\'as Cabrera 13-15, Universidad Aut\'onoma de Madrid, 28049 Madrid, Spain} \\
${}^3$ Departament d'Estructura i Constituents de la Mat\`eria and Institut de Ci\`encies del Cosmos,\\[-0.3em] 
Universitat de Barcelona, Mart\'{\i} i Franqu\`es, 08028 Barcelona, Spain\\ 
${}^4$ Departament de F\'{\i}sica Fonamental, Universitat de Barcelona, 08028 Barcelona, Spain\\[2mm] 

\vspace*{2.5cm}

\small{\bf Abstract} \\[5mm]
\end{center}
\begin{center}
\begin{minipage}[h]{16.0cm}
We consider 4d string compactifications in the presence of fluxes, and classify particles, strings and domain walls arising from wrapped branes which have charges conserved modulo an integer $p$, and whose annihilation is catalized by fluxes, through the Freed-Witten anomaly or its dual versions. The $\IZ_p$-valued strings and particles are associated to $\IZ_p$ discrete gauge symmetries, which we show are realized as discrete subgroups of 4d $U(1)$ symmetries broken by their Chern-Simons couplings to the background fluxes. We also describe examples where the discrete gauge symmetry group is actually non-Abelian. The $\IZ_p$-valued domain walls separate vacua which have different flux quanta, yet are actually equivalent by an integer shift of axion fields (or further string duality symmetries).  We argue that certain examples are related by T-duality to the realization of discrete gauge symmetries and $\IZ_p$ charges from torsion (co)homology. At a formal level, the groups classifying these discrete charges should correspond to a generalization of K-theory in the presence of general fluxes (and including fundamental strings and NS5-branes).
\end{minipage}
\end{center}
\newpage
\setcounter{page}{1}
\pagestyle{plain}
\renewcommand{\thefootnote}{\arabic{footnote}}
\setcounter{footnote}{0}

\tableofcontents

\vspace*{1cm}

\section{Introduction}

Flux compactifications in string theory have been the subject of extensive exploration (for reviews, see e.g. \cite{Ibanez:2012zz,Grana:2005jc,Douglas:2006es,Blumenhagen:2006ci}). Yet most studies about the flux-induced modifications of brane wrappings have focused on 4d spacetime filling branes (used for model building), and brane instantons. These results already signal interesting physics, including changes in supersymmetry conditions \cite{Martelli:2003ki,Cascales:2004qp,Gomis:2005wc,Martucci:2005ht,Martucci:2006ij,Martucci:2011dn}, and appearance of flux-induced supersymmetry breaking soft terms \cite{Grana:2002nq,Camara:2003ku,Grana:2003ek,Camara:2004jj}. But already at topological level, the presence of fluxes modifies the brane wrappings by the Freed-Witten (FW) consistency conditions (or their dual versions) \cite{Freed:1999vc,Maldacena:2001xj}. In this paper we consider wrapped branes with non-trivial topological interplay from fluxes, focusing on general 4d defects (strings, particles and domain walls), whose topological charges turn out to be conserved only modulo an integer $p$, related to the flux quanta.

For strings and particles, these $\IZ_p$-valued charges are related to the existence of $\IZ_p$ discrete gauge symmetries. This is similar to the realization of discrete gauge symmetries in terms of torsion homology classes in \cite{Camara:2011jg} (see also \cite{Gukov:1998kn}). An important difference is that we focus on branes wrapped on $\IZ$-valued homological cycles; these branes can however decay in sets of $p$ due to the effect of fluxes (dubbed `flux catalysis'). This is ultimately related to the fact that homology is in general not the right mathematical tool to classify brane charges. Already for D-brane charges must be classified by K-theory (in the absence of fluxes), or twisted K-theory (in the presence of NSNS 3-form flux)  \cite{Witten:1998cd}. In general, the mathematical tool classifying D-brane charges in the presence of RR fluxes (or including NS5-branes and fundamental strings) has not been determined. Our analysis can thus be regarded as a physical classification of stable objects and their decay processes in certain 4d flux compactifications (in analogy with the physical derivation of twisted K-theory in \cite{Maldacena:2001xj}), and hence a computation of the groups of brane charges, regardless of their underlying mathematical definition (see e.g. \cite{Evslin:2001cj,Evslin:2002sa,Evslin:2003hd,Evslin:2004vs,Collinucci:2006ug,Evslin:2007ti,Evslin:2007au} for  earlier discussions in this vein). In this respect, one relevant conclusion of our analysis is that these groups can be non-Abelian.  

The particles charged under these discrete symmetries are wrapped branes; although they are typically very massive, there are interesting setups in which they correspond to massless chiral fields, in particular in F-theory compactification on fourfolds. We describe the appearance of such symmetries and connect our discussion with recent developments in the physics of $U(1)$'s and brane instantons in F-theory, see e.g. \cite{Grimm:2010ez,Marsano:2011nn,Braun:2011zm,Kerstan:2012cy}.

We also encounter $\IZ_p$-valued domain walls, which can decay in sets of $p$ by nucleation of a string loop. Since domain walls from wrapped branes separate vacua differing in their background fluxes, our analysis provides a physical derivation of certain $\IZ_p$-valued flux quantization conditions. This goes beyond the naive characterization of fluxes as integer cohomology classes, and is related to the fact that integer cohomology is not the right mathematical tool to classify $p$-form fluxes in string theory (see  \cite{Moore:1999gb} for a partially more complete definition, in terms of K-theory classes). We also show that $\IZ_p$-valued domain walls can be associated to (infinite) discrete gauge symmetries, corresponding to non-trivial (axion-like) field identifications in the theory (and ultimately to string dualities), spontaneously broken by the presence of the background fluxes.

We focus on type II compactifications (with background fluxes). Some aspects of $\IZ_p$ strings in compactifications with fluxes have been considered in \cite{Copeland:2003bj} for type IIB (with NSNS and RR 3-form flux) and in \cite{Polchinski:2005bg} for the heterotic (with gauge fluxes). Also, we focus on compactifications with $p$-form fluxes, and do not introduce geometric or non-geometric fluxes (although the former are included, in the sense that the geometries need not restrict to the Calabi-Yau (CY) case, because supersymmetry is not essential to the analysis). By duality, such fluxes will lead to similar effects, or suitable generalizations.

The paper is organized as follows. In section \ref{ch2} we show that background fluxes induce the appearance of 4d discrete gauge symmetries, and use them to characterize $\IZ_p$-valued particles and strings in 4d compactifications with fluxes. This is applied in section \ref{sec:fth} to F/M-theory compactifications with $G_4$-flux, which provides a setup in which some of the $\IZ_p$ charged particles can be massless, and relate the analysis to recent developments in F/M-theory $U(1)$'s and M5-brane instantons. In section \ref{sec:non-abelian} we present a realization of non-Abelian discrete gauge symmetries  induced by fluxes, on geometries admitting non-trivial 1-cycles. In section \ref{sec:combine} we study the combination of different fluxes, and show that the combination of NSNS and RR fluxes often required for 4d Minkowski vacua, introduce incompatible gaugings in the 4d theory, which are rendered compatible by extra ingredients, like extra (anti-)branes, or orientifold planes; the latter project out the discrete gauge symmetries, and hence the corresponding particles and strings. In section \ref{sec:udw} we turn to $\IZ_p$-valued domain walls, and their relation to string duality symmetries relating vacua with different flux quanta. In section \ref{sec:torsion-hom} we use dualities to connect the flux-induced $\IZ_p$ charges to those appearing in compactification with torsion (co)homology classes. In section \ref{sec:conclu} we offer some final remarks. To make the discussion self-contained, appendix \ref{sec:field-th} reviews $\IZ_p$ gauge symmetries, and appendix  \ref{fw-hw} reviews the Hanany-Witten and Freed-Witten like effects appearing in the text. Appendix \ref{sec:isometries} provides a brief discussion of flux-induced discrete gauge symmetries, including a non-Abelian example, arising from Kaluza-Klein (KK) $U(1)$'s in compactifications with isometries.

\section{Discrete gauge symmetries from flux catalysis}
\label{ch2}

\subsection{Generalities and flux catalysis}

We start the discussion considering wrapped branes that give rise to particles and strings with $\IZ_p$-valued charges. These signal the existence of discrete gauge symmetries, which interestingly can be identified by the presence of $BF$ couplings in the 4d theory, see appendix \ref{sec:field-th}.  In type II models  
these couplings arise from KK reduction of the 10d  Chern-Simons couplings, which are of the form
\beqa
(a) \quad \int_{10d} H_3\wedge F_p\wedge C_{7-p} \quad , \quad 
\label{cs1} 
(b) \quad \int_{10d} B_2\wedge F_p\wedge F_{8-p} \label{cs2}
\eeqa
Here $B_2$ and $H_3$ denote the NSNS 2-form potential and its field strength, whereas $C_n$ and $F_{n+1}$ denote the RR $n$-form potential and its field strength, with $n$ even or odd for type IIA or IIB theories, respectively (and including the 0-form field strength $F_0$, namely massive IIA theory). Although the above two expressions are locally equivalent upon integration by parts, we keep both for convenience. 

In a compactification with non-trivial fluxes the above couplings lead to $B\wedge F$ terms in the 4d theory describing $\IZ_p$ discrete gauge symmetries. As discussed in \cite{Banks:2010zn} and reviewed in Appendix \ref{sec:field-th}, a 4d 2-form field $B_2$ with a coupling $p\, B_2\wedge F$ to a $U(1)$ gauge field (in the normalization that the dual scalar has period 1 and the minimal electric charge is unity) can be dualized to a 4d scalar $\phi$ breaking spontaneously the $U(1)$ down to a $\IZ_p$ subgroup. Morally, the scalar $\phi$ is analogous to the phase of a charge $p$ scalar whose modulus acquires a vev.

The $\IZ_p$ symmetry will prove a useful tool to classify branes in the system. The $\IZ_p$-valued particles and strings charged under the discrete gauge symmetry are given by branes wrapped on homologically non-trivial $\IZ$-valued cycles, yet they can decay in sets of $p$, due to processes allowed by the presence of fluxes (dubbed `flux catalysis'). Prototypes of such processes are the decay of D-branes ending on a higher-dimensional brane with non-trivial NSNS flux along its worldvolume \cite{Maldacena:2001xj}, due to the Freed-Witten consistency condition\footnote{Actually \cite{Freed:1999vc} considered the case of torsion $H_3$, and the physical picture for general $H_3$ appeared in \cite{Maldacena:2001xj}. Still, we stick to the widely used term FW anomaly / consistency condition, even for non-torsion $H_3$.},  or the decay of fundamental strings on a D$p$-brane with non-trivial $p$-form flux along its worldvolume (as in the baryon vertex in \cite{Witten:1998xy}). 

One may think that the charge carried by the instantonic object sources a flux which is left behind after the decay, such that the $p$ charged objects do not decay to the vacuum. We emphasize that in the present 4d case the decay of $p$ charged objects is to the vacuum of the theory, with no flux left behind.\footnote{Nevertheless, the decay can lead to short lived fluxes, whose small range of interaction is controlled by the scale of the massive gauge boson, and which do not modify the topological characterization of the charges.} This is particularly clear in the language of 4d field theory, where the field theory operator describing the instantonic decays have no long range effect \cite{Banks:2010zn}, as we describe in Appendix \ref{sec:field-th}. We do not consider other setups, in other spacetime dimensions (actually, other codimensionalities of the $\IZ_k$ charged branes), where there is some left-over flux which demands a combined classfication of D-branes and fluxes \cite{Evslin:2003hd}.

In this section we scan through different possibilities of background fluxes, and the resulting discrete gauge symmetries and $\IZ_p$-valued wrapped branes, for the case of compactifications with NSNS and RR fluxes. We make some simplifying assumptions, some of which will be relaxed in the following sections. In particular, we take the internal manifold to be CY. This simplifies the flux wrapping possibilities, since $b_1=b_5=0$. Generalization is possible, and can lead to richer structures, like the non-Abelian discrete symmetries in section \ref{sec:non-abelian}. Moreover, we do not include geometric or non-geometric fluxes, beyond some comments in section \ref{sec:torsion-hom}. Consequently, the presence of the NSNS flux makes the discussion not invariant under mirror symmetry, and must be carried out in IIA and IIB models independently. Finally, we mainly consider compactifications with only one kind of flux. This simplifies the discussion and also allows to postpone the introduction of orientifold planes.
These are naturally discussed in section \ref{sec:combine}, which treats certain combinations of fluxes. Other combined flux configurations, leading to non-Abelian discrete symmetries, are considered in section \ref{sec:non-abelian} and appendix \ref{sec:isometries}. 

Hence, for type IIA, we consider turning on 0-form flux (subsection \ref{sec:IIAform0}), 2-form flux (subsection \ref{sec:IIAform2}), NSNS 3-form flux (subsection \ref{sec:IIAform3}), 4-form flux (subsection \ref{sec:IIAform4}) or 6-form flux (subsection \ref{sec:IIAform6}). Our ordering of sections does not follow the degree, in order to postpone some more involved symmetries to the end. For type IIB we consider turning on NSNS or RR 3-form flux (subsection \ref{sec:IIB}). Since the logic of the derivation is similar for each type of flux, some parts of this section might seem slightly repetitive. We summarize  the main results of this section in tables \ref{iiatable} and \ref{iibtable}, so that readers can use them for quick reference.

\subsection{Massive IIA}
\label{sec:IIAform0}

Consider massive type IIA string theory with  mass parameter ${\ov F}_0=p$, compactified on a CY $\IX_6$, where here and in what follows, overlining indicates the background value. The Chern-Simons couplings (\ref{cs2}b) contain a term producing a 4d $BF$ coupling
\beqa
\int_{10d} {\ov F}_0\, B_2\wedge F_8\quad \to\quad p\int_{4d} B_2\wedge \hat F_2
\label{bf-massive-iia}
\eeqa
with $\hat F_2\, =\,\int_{\IX_6} F_8$. The theory automatically has a $\IZ_p$ discrete gauge symmetry, for which the $\IZ_p$-charged particles are the D6-branes wrapped on $\IX_6$, and the $\IZ_p$-charged strings are the fundamental strings (F1s), see appendix \ref{sec:field-th}.

The instanton annihilating $p$ (minimally) charged particles is the object coupling to the scalar 4d-dual to $B_2$, namely an NS5-brane wrapped on $\IX_6$. Microscopically, the presence of a mass parameter $p$ induces a worldvolume tadpole on the NS5-brane,  which must be cancelled by the addition of $p$ D6-branes ending on the NS5-brane \cite{Hanany:1997gh}, see appendix \ref{fw-hw}. Similarly, charged strings can annihilate in sets of $p$ in a junction given by the object charged magnetically under $C_7$, namely a D0-brane. Indeed, a D0-brane in the presence of a mass parameter $p$ must be dressed with $p$ fundamental strings\footnote{The fact that D0-branes are not states in the theory dovetails the absence of an 11d M-theory lift of massive IIA theory, see \cite{Aharony:2010af} for a recent discussion.}  \cite{Bergman:1997py}.

As noted in the introduction, the F1s and the wrapped D6-branes have physical $\IZ_p$-valued charges, in contrast with their naive $\IZ$-valued nature in homology and even in K-theory. Similar remarks apply to the diverse $\IZ_p$ branes in the coming subsections.
 
\subsection{Type IIA Freund-Rubin}
\label{sec:IIAform6}

Let us now instead consider type IIA compactifications on $\IX_6$ with $p$ units of ${\ov F}_6$ flux on it (for instance, as in the Freund-Rubin compactifications, ubiquitous in the AdS$_4$/CFT$_3$ correspondences initiated by \cite{Aharony:2008ug}). The 10d CS couplings (\ref{cs2}b) lead to a 4d $BF$ coupling as follows
\beqa
\int_{10d} B_2\wedge F_2\wedge {\ov F}_6\,\to\; p\int_{4d} \, B_2\wedge F_2
\eeqa
signalling a $\IZ_p$ symmetry. Charged particles are D0-branes, which annihilate in units of $p$ on an instanton, given by an NS5-brane wrapped on $\IX_6$. Charged strings are fundamental strings, which annihilate in units of $p$ on a D6-brane wrapped on $\IX_6$ (this is analogous to the 
AdS$_4$/CFT$_3$ version of the baryon in \cite{Witten:1998xy}). Morally, this system relates via six  `T-dualities' (or equivalently a 4d electric-magnetic transformation) to that of the previous section.
 
\subsection{Type IIA with NSNS flux}
\label{sec:IIAform3}

For the case of type IIA compactifications with NSNS 3-form flux ${\ov H}_3$ we introduce  a symplectic basis of 3-cycles $\{\alpha_k\}$, $\{\beta_k\}$, with $\alpha_k\cdot \beta_l=\delta_{kl}$, and define
\beqa
&\int_{\a_k} {\ov H}_3=p_k\quad , &\quad \int_{\b_k} {\ov H}_3=p_k' \nonumber \\
&\int_{\b_k} C_5\, =\hat B_{k} \quad ,& \quad \int_{\a_k} C_5=\hat B_k'
\eeqa
There are 4d $BF$ couplings from the reduction of the 10d Chern-Simons term (\ref{cs1}a): 
\beqa
\int_{10d} {\ov H}_3\wedge F_2 \wedge C_5 \,\to\,  \sum_k \int_{4d} (\,p_k \, \hat B_k\, -\,p_k'\,\hat B_k'\,) \wedge F_2
\eeqa
where $F_2$ is the field strength of the RR 1-form. This $U(1)$ group is broken to a remnant $\IZ_p$ discrete gauge symmetry, with  $p={\rm gcd}(p_k,p_k')$.

Particles charged under this symmetry are D0-branes, which can annihilate in sets of $p_k$ (resp. $p_k'$) by D2-brane instantons wrapped on $\a_k$ (resp. $\b_k$), as dictated by the Freed-Witten anomaly induced by ${\ov H}_3$. In particular, using Bezout's lemma, there exist integers $n_k$ and $m_k$ such that $\sum_k(n_kp_k+m_kp_k)=p$, so that a D2-brane on $\sum_k (n_k\a_k +m_k\b_k)$ annihilates exactly $p$ D0-branes. 

Similarly, the basic charged string is  a D4-brane on the linear combination 3-cycle\footnote{D4-branes wrapped on linear combination 3-cycles with FW inconsistencies have D2-branes attached, and correspond to the strings bounding domain walls discussed section \ref{sec:udw}.}
\beqa
\Pi=\sum_k\,\Big(\, \frac{p_k}{p}\,\b_k - \frac{p_k'}{p}\,\a_k\, \Big)
\eeqa 
which is free from FW inconsistencies, $\int_{\Pi} H_3=0$. 
The junction annihilating $p$ charged strings is a D6-brane wrapped on $\IX_6$; due to its Freed-Witten anomaly it emits D4-branes in the total class $\sum_k\left(p_k[\b_k]-p_k'[\a_k]\right)$, namely $p$ minimally charged strings.

\subsection{Type IIA with $F_2$ flux}
\label{sec:IIAform2}

In this case it is convenient to introduce a basis of 2-cycles $\{\Pi_k\}$ and dual 4-cycles $\{\Gamma_l\}$ (with $\Pi_k\cdot \Gamma_l=\delta_{kl}$) and define
\beqa
\int_{\Pi_k} {\ov F}_2= p_k\quad , \quad \int_{\Gamma_l} F_6\, =\, \hat F_2^l
\label{f2-int}
\eeqa
Then a 10d Chern-Simons coupling (\ref{cs2}b) descends to a 4d $BF$ coupling as follows
\beqa
\int_{10d} B_2\wedge {\ov F}_2\wedge F_6 \,\to\, \sum_k\int_{4d} p_k \, B_2\wedge \hat F_2^k
\eeqa
The remnant discrete gauge symmetry is not manifest by inspection. Superficially, it may seem that each $U(1)$ factor  leaves an unbroken $\IZ_{p_k}$. This is however not correct, since the different $U(1)$ factors couple simultaneously to a {\em single} 2-form field. Indeed, there is an unbroken $U(1)^{N-1}$, with generators $Q_a$, $a=1,\ldots , N-1$, given by integer linear combinations 
\beqa
Q_a=\sum_k c_k^a\, Q_k\quad , \quad c_k^a\in\IZ
\label{comb-uunos}
\eeqa
with $\vec{c}_a\cdot \vec{p}=0$.  The only broken $U(1)$ linear combination is the one orthogonal to all the massless $U(1)$'s, namely it is given by $Q=\sum_k (p_k/p) Q_k$, where the factor  $p={\rm gcd}(p_k)$ is included to keep the normalization such that minimal charge is 1, see appendix \ref{sec:field-th}. Its $BF$ couplings are
\beqa
\sum_k\frac{(p_k)^2}{p} \,B_2\wedge F_2
\label{tricky-symm}
\eeqa
The symmetry is therefore $\IZ_q$ with $q=\sum_k (p_k)^2/p$. 

Particles charged under this symmetry are D4-branes wrapped on 4-cycles\footnote{The minimal charge particle is a D4-brane on $\sum_k n_k \Gamma_k$ with integers $n_k$ satisfying $\sum_k n_k p_k=p$, which exists by Bezout's lemma.}, and the relevant instanton that annihilates them is an NS5-brane wrapped on $\IX_6$. Indeed, due to the FW anomaly induced by ${\ov F}_2$, the NS5 must emit sets of $p_k$ D4-branes wrapped on $\Gamma_k$,  for all $k$. Since each D4$_k$-brane particle has charge $+1$ under $Q_k$,  the charge violation for a combination $Q_a$ in (\ref{comb-uunos}) is $\sum_k c^a_k\, p_k$. This vanishes for massless $U(1)$'s, whereas $Q$ is violated in $q$ units. 

Charged strings are fundamental strings, and their $\IZ_q$-valued nature is not automatically manifest. Indeed, F1s can annihilate in sets of $p_k$ on a D2-brane on $\Pi_k$, naively violating the $\IZ_q$ symmetry (as $p_k$ is in general not a multiple o $q$). However such D2-brane is not gauge invariant, since it also carries monopole charge under some unbroken $U(1)$'s (concretely, any linear combination involving $Q_k$). The actual gauge invariant junction annihilating F1s is a D2-brane on the 2-cycle $\Pi=\sum_k \frac{p_k}{p}\Pi_k$, which has no monopole charge and emits $q$ F1s, in agreement with the $\IZ_q$ symmetry. 

\subsection{Type IIA with $F_4$ flux}
\label{sec:IIAform4}

We finish the discussion of type IIA fluxes by considering compactifications with ${\ov F}_4$ flux for the field strength of the RR 3-form. This is morally related to the system in the previous section by six `T-dualities',  so the discussion is similar. We introduce dual 2-cycles $\{\Pi_k\}$ and 4-cycles $\{\Gamma_l\}$ and define
\beqa
\int_{\Gamma_k} {\ov F}_4= p_k\quad , \quad \int_{\Pi_k} F_4\, =\, \hat F_2^k
\eeqa
Then a 10d Chern-Simons coupling (\ref{cs2}b) descends to a 4d $BF$ coupling as follows
\beqa
\int_{10d} B_2\wedge {\ov F}_4\wedge F_4 \,\to\, \sum_k \int_{4d} p_k \,B_2\wedge \hat F_2^k
\eeqa
leading to the same discrete symmetry pattern as in section \ref{sec:IIAform2}. In short, the charged particles are D2-branes on 2-cycles, and annihilate on an NS5-brane wrapped on $\IX_6$, and the charged strings are fundamental strings, and annihilate on a D4-brane on the linear combination 4-cycle $\sum_k \frac{p_k}{p}\Gamma_k$.

\subsection{Type IIB with 3-form flux}
\label{sec:IIB}

A popular class of flux compactifications is that of type IIB on a CY $\IX_6$ with NSNS and RR 3-form fluxes \cite{Dasgupta:1999ss,Giddings:2001yu}. As explained, we momentarily consider introducing a single kind of flux, say RR 3-form flux, and postpone the combination of fluxes to section \ref{sec:combine}.

Thus, let us first consider switching only on RR 3-form flux.  Without loss of generality, we can introduce an adapted symplectic basis of 3-cycles $\{\alpha_k\}$, $\{\beta_k\}$, such that there is flux only on the $\alpha$ cycles, and define
\beqa
 \int_{\alpha_k} {\ov F}_3=p_k\quad , \quad  \int_{\beta_k} F_5\, =\hat F_2^{k}
\eeqa
The 10d CS coupling leads to the 4d $BF$ terms
\beqa
\int_{10d} {\ov F}_3\wedge B_2 \wedge F_5 \,\to\,  \sum_k \int_{4d}  p_k\, B_2\wedge  \hat F_2^k
\eeqa
In analogy with section \ref{sec:IIAform2}, we denote $p={\rm gcd}(p_k)$, and the discrete symmetry is $\IZ_q$ with $q=\sum_k \frac{(p_k)^2}{p}$. Charged particles are wrapped D3-branes, and annihilate on an NS5-brane on $\IX_6$, while charged strings are fundamental strings, and annihilate on a D3-brane in a linear combination 3-cycle.

A similar setup is obtained for compactification with only NSNS 3-form flux, with the replacements $F_3\to H_3$ and $B_2\to C_2$, as expected from S-duality. For the resulting discrete symmetry, the charged particles are wrapped D3-branes, and annihilate on a D5-brane, while charged charged strings are D1-branes, and annihilate on a D3-brane wrapping a linear combination 3-cycle.

Clearly, acting with transformations in the 10d $SL(2,\IZ)$ duality group, we can obtain configurations with combined NSNS and RR 3-form fluxes, albeit a restricted class. In particular, since the starting configurations have zero contribution to the D3-brane tadpole $\int_{\IX_6} {\ov F}_3\wedge {\ov H}_3$, and this is $SL(2,\IZ)$ invariant, this strategy cannot reach configurations with non-zero D3-brane tadpole. General flux configurations, and the role of the tadpole will be discussed in section \ref{sec:combine}.\\

\begin{table}[!ht]
\begin{center}
{\begin{tabular}{|c|c|c||c|c||c|c||c|c||c|c|}
\hline
$p$ & $\hat B_2$ & $\hat F_2$ & \multicolumn{2}{|c||}{\bf Particle}&\multicolumn{2}{|c||}{\bf Instanton} & \multicolumn{2}{|c||}{\bf String} & \multicolumn{2}{|c|}{\bf Junction}\\
\hline
& & & type & cycle & type & cycle &  type & cycle &  type & cycle \\
\hline\hline
${\ov F}_0$ & $B_2$ & $\int_{{\bf X}_6}F_8$ & D6 & ${\bf X}_6$ & NS5 & ${\bf X}_6$ & F1 & $-$ & D0 & $-$\\
\hline
$\int_{\Pi_2}{\ov F}_2$ & $B_2$ & $\int_{\Gamma_4}F_6$ & D4 & $[\Gamma_4]$ & NS5 & ${\bf X}_6$ & F1 & $-$ & D2 & $[\Pi_2]$\\
\hline
$\int_{\alpha_3}{\ov H}_3$ & $\int_{\beta_3}C_5$ & $F_2$ & D0 & $-$ & D2 & $[\alpha_3]$ & D4 & $[\beta_3]$ & D6 & ${\bf X}_6$\\
\hline \hline
$\int_{{\bf X}_6}{\ov F}_6$ & $B_2$ & $F_2$ & D0 & $-$ & NS5 & ${\bf X}_6$ & F1 & $-$ & D6 & ${\bf X}_6$\\
\hline
$\int_{\Gamma_4}{\ov F}_4$ & $B_2$ & $\int_{\Pi_2}F_4$ & D2 & $[\Pi_2]$ & NS5 & ${\bf X}_6$ & F1 & $-$ & D4 & $[\Gamma_4]$\\
\hline
$\int_{\beta_3}{\ov H}_3$ & $\int_{\alpha_3}C_5$ & $F_2$ & D0 & $-$ & D2 & $[\beta_3]$ & D4 & $[\alpha_3]$ & D6 & ${\bf X}_6$\\
\hline
\end{tabular}}
\caption{Summary of 4d St\"uckelberg couplings $p\, \hat B_2\wedge \hat F_2$ induced by single background flux components in type IIA CY compactifications. We also present the higher-dimensional origin of the charged particles and strings, together with that of the instantons and junctions that respectively anhihilate them. The notation is such that $[\Pi_2]\in H_2({\bf X}_6,\IZ)$, $[\Gamma_4]\in H_4({\bf X}_6,\IZ)$, $[\alpha_3],[\beta_3]\in H_3({\bf X}_6,\IZ)$ and $[\Pi_2]\cdot[\Gamma_4]=[\alpha_3]\cdot[\beta_3]=1$. The upper and lower halves of the table are related by a 4d electric-magnetic transformation.\label{iiatable}}
\end{center}
\end{table}

\begin{table}[!ht]
\begin{center}
{\begin{tabular}{|c|c|c||c|c||c|c||c|c||c|c|}
\hline
$p$ & $\hat B_2$ & $\hat F_2$ & \multicolumn{2}{|c||}{\bf Particle}&\multicolumn{2}{|c||}{\bf Instanton} & \multicolumn{2}{|c||}{\bf String} & \multicolumn{2}{|c|}{\bf Junction}\\
\hline
& & & type & cycle & type & cycle &  type & cycle &  type & cycle \\
\hline\hline
$\int_{\alpha_3}{\ov F}_3$ & $B_2$ & $\int_{\beta_3}F_5$ & D3 & $[\beta_3]$ & NS5 & ${\bf X}_6$ & F1 & $-$ & D3 & $[\alpha_3]$\\
\hline
$\int_{\alpha_3}{\ov H}_3$ & $C_2$ & $\int_{\beta_3}F_5$ & D3 & $[\beta_3]$ & D5 & ${\bf X}_6$ & D1 & $-$ & D3 & $[\alpha_3]$\\
\hline \hline
$\int_{\beta_3}{\ov F}_3$ & $B_2$ & $\int_{\alpha_3}F_5$ & D3 & $[\alpha_3]$ & NS5 & ${\bf X}_6$ & F1 & $-$ & D3 & $[\beta_3]$\\
\hline
$\int_{\beta_3}{\ov H}_3$ & $C_2$ & $\int_{\alpha_3}F_5$ & D3 & $[\alpha_3]$ & D5 & ${\bf X}_6$ & D1 & $-$ & D3 & $[\beta_3]$\\
\hline
\end{tabular}}
\caption{Summary of 4d St\"uckelberg couplings $p\, \hat B_2\wedge \hat F_2$ induced by single background flux components in type IIB CY compactifications. We also present the higher-dimensional origin of the charged particles and strings, together with that of the instantons and junctions that respectively anhihilate them. The notation is such that $[\alpha_3],[\beta_3]\in H_3({\bf X}_6,\IZ)$ and $[\alpha_3]\cdot[\beta_3]=1$. The upper and lower halves of the table are related by a 4d electric-magnetic transformation.\label{iibtable}}
\end{center}
\end{table}

\section{$G_4$-flux catalysis in F-theory}
\label{sec:fth}

The examples in the previous section seem of mere academic interest, since the charged particles arise from wrapped branes which are typically very massive, so the symmetries seem to act trivially on the massless spectrum. However, there are situations where also the massless chiral spectrum arises from wrapped branes, the prototypical example being F/M-theory on a CY fourfold with $G_4$-flux \cite{Donagi:2008ca,Beasley:2008dc,Beasley:2008kw,Donagi:2008kj}. In this section we describe the appearance of discrete gauge symmetries from $G_4$-flux catalysis in F-theory. The result reproduces the perturbative D-brane description in \cite{BerasaluceGonzalez:2011wy}, and also dovetails the discussion in \cite{Kerstan:2012cy} about instantons in F/M-theory and their violation of $U(1)$ symmetries. It also clarifies the role of M5-brane instantons which are not vertical and therefore do not survive as Euclidean D3-brane instantons in the F-theory limit \cite{Witten:1996bn}. The punchline is that they become line operators that annihilate charged strings.

We consider  F/M-theory on an elliptically fibered Calabi-Yau fourfold $\IX_8$ over a base ${\bf B}_3$, in the presence of non-trivial $G_4$ flux. The latter is usually discussed in the context of  M-theory on the resolved fourfold ${\hat\IX}_8$, in which the singularity structure yielding the gauge group on the 7-branes has been smoothed out.
The singularities associated with the non-Abelian piece $G'$ of the total gauge group are resolved by introducing a set of ${\rm rk}(G')$ resolution divisors $E_i$. The realization of extra $U(1)$ gauge symmetries has been argued in  \cite{Grimm:2010ez} to yield an extra set of resolution divisors $S_A$. We can denote these divisors collectively by $S_\Lambda$, and their Poincar\'e dual 2-forms $\omega_\Lambda$. These 2-forms give rise to 3d gauge bosons $A_\Lambda$ from the KK expansion
\beqa
C_3=A_\Lambda\wedge \omega_\Lambda+\ldots
\eeqa
The introduction of a flux background $G_4$ in M-theory on ${\hat\IX}_8$ leads to 4d $BF$ couplings and discrete gauge symmetries in F-theory, analogously to what has been discussed in earlier type II examples. A difference, however, is that they appear as 3d $A_1\wedge \hat F_2$ couplings between a 1-form potential $A_1$ (lifting to the 4d 2-form in F-theory) and the $U(1)$ field strength $\hat F_2$. The couplings arise as usual from the 11d CS coupling
\beqa
\int_{11d} C_3\wedge {\ov G}_4\wedge G_4 \, \to \, \int_{3d} \Theta_{\Lambda\Gamma}\, A_\Lambda \wedge \hat F_\Gamma
\eeqa 
with
\beqa
\Theta_{\Lambda\Gamma}=2\int\omega_\Lambda\wedge \omega_\Gamma\wedge {\ov G}_4
\label{fth-coeff}
\eeqa
Considering linear combinations $\sum_\Gamma c_\Gamma Q_\Gamma$, with $c_\Gamma\in\IZ$ and  $\textrm{gcd}(c_\Gamma)=1$, and defining $p_\Lambda=\sum_\Gamma c_\Gamma\Theta_{\Lambda\Gamma}$, there is an unbroken $\IZ_p$ subgroup with $p={\rm gcd}(p_\Lambda)$. 

Fluxes ${\ov G}_4$ that lift to F-theory morally carry `one leg along the elliptic fiber', and three along the base ${\bf B}_3$. Hence, in (\ref{fth-coeff}) they couple forms $\omega_\Lambda$ not on ${\bf B}_3$ (namely, with one leg along the elliptic fiber) with forms $\omega_\Sigma$ entirely on ${\bf B}_3$.
Charged $\IZ_p$ particles arise from M2-branes on 2-cycles $\Pi\not\in{\bf B}_3$ (namely, with one leg along the elliptic fiber), and carry $\IZ_p$ charge $\sum_\Gamma c_\Gamma\int_\Pi \omega_\Gamma$ (mod $p$). Instantons arise from M5-branes on vertical divisors \cite{Witten:1996bn}\footnote{Namely, on 6-cycles that result from fibering the $\IT^2$ fiber over a divisor on the base ${\bf B}_3$, on which the dual D3-brane instanton wraps in F-theory.}, and emit M2-branes due to a FW anomaly  \cite{Kerstan:2012cy} in multiples of $p$. The 4d charged strings arise in the 3d theory as charged particle-like vortices, coupled to $A_\Lambda$. They arise from M2-branes wrapped on the 2-cycles of the base ${\bf B}_3$, dual to $\omega_\Lambda$. The 4d junction annihilating charged strings corresponds to an instanton in the 3d theory annihilating vortices, arising from an M5-brane wrapped on the divisor $S_\Lambda$ (which has a ${\ov G}_4$ FW anomaly as required). Note that the latter is not a vertical divisor, hence does not turn into a 4d instanton but rather into a junction in the dual F-theory setup.

As a simple F-theory example illustrating the existence of massless particles charged under the $\IZ_p$ symmetries, consider a flux along a non-Abelian generator, breaking $G'=SU(N)\to SU(N_1)\times SU(N_2)\times U(1)$. Denoting by $E_{i_0}$ and $\omega_{i_0}$ respectively the divisor and 2-form associated to this $U(1)\in G'$, we have
\beqa
{\ov G}_4\, =\, {\ov F}_2 \wedge \omega_{i_0}\quad , \quad {\rm with}\; {\ov F}_2\in H^{(1,1)}({\bf B}_3,\IZ)
\eeqa
In IIB language, we have a stack of 7-branes supporting a $SU(N)$ gauge group on a 4-cycle $S\in {\bf B}_3$, broken by a worldvolume flux ${\ov F}_2$ (in this case, non-trivial in ${\bf B}_3$). The localization over $S$ of the forms $\omega_i$ associated to generators of $G'$  is encoded in the relation 
\beqa
\omega_i\wedge \omega_j=C_{ij} \delta_2(S)
\eeqa
where $C_{ij}$ is the Cartan matrix of $SU(N)$ and $\delta_2(S)$ is the Poincar\'e dual of $S$ in ${\bf B}_3$. The 3d $AF$ couplings for the $U(1)$ are
\beqa
\int_{3d} p_\Lambda A_\Lambda \wedge F_2\quad ,\quad {\rm with}\, \; p_\Lambda=\int_{\IX_8} \omega_{\Lambda}\wedge \omega_{i_0} \wedge {\ov F}_2 \wedge \omega_{i_0}=\int_{S} \omega_\Lambda\wedge {\ov F}_2
\eeqa
The corresponding F-theory 4d $BF$ coupling agrees with that arising from the 8d CS on the 7-brane worldvolume, $\int_{8d}C_4\wedge {\ov F}_2\wedge F_2$. The particles arising from M2-branes on the 2-cycles dual to the $E_i$, associated to generators of $G'=SU(N)$ broken by the flux, are states transforming in bi-fundamental representations $(N_1,{\ov N}_2)$ and charged under $\IZ_p$. In F-theory, these states are massless 8d particles, and can lead to massless 4d particles (if there are suitable zero modes in the KK reduction on $S$).

It is also possible to provide examples of massless charged particles for $\IZ_p$ symmetries that arise from $U(1)$'s not in $G'$, see \cite{BerasaluceGonzalez:2011wy} for the discussion of discrete remnants of $U(1)_{B-L}$ in the MSSM-like F-theory models of \cite{Blumenhagen:2009yv}.

\section{Non-Abelian discrete gauge symmetry from fluxes}
\label{sec:non-abelian}

When fluxes coexist, there can be several $\IZ_p$ factors, which may be non-commuting, resulting in a non-Abelian discrete gauge symmetry\footnote{See \cite{Alford:1989ch,Alford:1990mk,Alford:1990pt,Alford:1991vr,Alford:1992yx,Lee:1994qg} for early field theory literature, and \cite{Gukov:1998kn,BerasaluceGonzalez:2012vb} for string realizations in type II and \cite{Kobayashi:2006wq,Nilles:2012cy} in heterotic orbifolds.}, as we show in a class of examples in this section. In field theoretical grounds, the non-Abelianity implies that strings associated to non-commuting elements $g$ and $h$, when crossing each other, produce a new string stretching between them, associated to the commutator $c=ghg^{-1}h^{-1}$. Microscopically, these strings appear by the Hanany-Witten brane creation effects \cite{Hanany:1996ie} discussed in appendix \ref{fw-hw}. Note that although the strings created in the crossing are finite in extent, the theory must also contain stable infinite strings associated to the generators $c$, which therefore describe genuine elements of the discrete symmetry group of theory. The string creation effect is however an excellent tool to diagnose the non-Abelianity of the discrete gauge symmetry in a theory.

Non-Abelian discrete symmetries and brane/string creation effects have been realized in cases with discrete symmetries that arise from $p$-forms on torsion cohomology classes in \cite{Gukov:1998kn,BerasaluceGonzalez:2012vb}.
In this section we show that non-Abelianity for discrete symmetries can also arise from flux catalysis. We consider models with two $\IZ_p$ symmetries, with charged strings given by D4-branes on two 3-cycles on $\IX_6$, which upon crossing produce fundamental strings sitting at the 3-cycle intersection points in $\IX_6$, and stretching in 4d\footnote{Or similarly for other D-branes intersecting at points in $\IX_6$, e.g. D3- and D5-branes on 2- and 4-cycles, respectively. Also, in appendix \ref{sec:isom-nonab} we get non-Abelianity from the HW effect for NS5-branes.}.

Wrapped D4-branes playing the role of $\IZ_p$-charged strings have appeared in section \ref{sec:IIAform3}. However, the structure of the Chern-Simons terms is such that it leads to the gauging of only {\em one} $U(1)$, and cannot accommodate two independent discrete generators. To overcome this point, we consider geometries with non-trivial 1-cycles, which allow a richer set of gauge bosons and Chern-Simons couplings. Such geometries include not only $\IT^6$ and K3$\times\IT^2$, but more general cases beyond the CY realm, for instance certain twisted tori, or general $\IT^2$ bundles\footnote{These are not fibrations since the 1-cycles do not pinch at any point on the base.} over a base $\IB_4$ (such as those appeared in \cite{Becker:2008rc}  in the heterotic context). For simplicity, we will have in mind a product $\IB_4\times \IT^2$, keeping $\IB_4$ arbitrary, since we are not particularly interested in supersymmetry. Note that it should be possible to obtain non-Abelian discrete symmetries without resorting to 1-cycles if one admits extra sources of gauging, like torsion homology classes or the presence of additional fluxes (e.g. geometric or non-geometric). We however stick to pure flux catalysis, and allow for 1-cycles, leaving other possibilities for future work.

Consider $\IX_6=\IB_4\times \IT^2$, and introduce the $\IT^2$ 1-cycles $a$ and $b$, and two dual basis $\{\Pi_k\}$ and $\{\Pi_k'\}$ of 2-cycles in $\IB_4$ (namely $\Pi_k\cdot\Pi_l'=\delta_{kl}$). We introduce fluxes for the RR 2-form field strength, and define the 4d forms and flux quanta
\beqa
 \int_{\Pi_k}{\ov F}_2\, =\, p_k \quad ,\quad &
 \quad \int_{a}H_3\, =\, \hat F_2^a \quad , & \quad  \int_{b}H_3\, =\, \hat F_2^b \nonumber \\
\int_{\Pi_k'\times b} C_5\, =\,  \hat B_k \quad , & \quad \int_{\Pi_k'\times a} C_5\, =\, \hat B_k'
\eeqa
The 10d Chern-Simons term (\ref{cs1}) descends in the presence of this backrgound to 4d $BF$ couplings as
\beqa
\int_{10d}{\ov F}_2\wedge H_3\wedge C_5\quad \to\quad \int_{4d} \bigg(\, \sum_k\, p_k\, \hat B_k\wedge \hat F_2^a \, -\, \sum_k \,p_k\, \hat B_k'\wedge \hat F_2^b\, \bigg)
\eeqa
Defining $p={\rm gcd}(p_k)$, each of the two $U(1)$ gauge factors is broken to a $\IZ_p$ gauge symmetry. The particles that are charged under this symmetry are F1s winding around the $a$ and $b$ 1-cycles, and annihilate on D2-brane instantons on the 3-cycles $\Pi_k'\times b$ and $\Pi_k'\times a$. The charged strings are D4-branes on the 3-cycles $\Delta=\sum_k (p_k/p) \Pi_k'\times b$ and $\Delta'=\sum_k (p_k'/p') \Pi_k'\times a$, and annihilate on junctions from NS5-branes on the 5-cycles $b\times \IB_4$ or $a\times \IB_4$. 

Crossing two 4d strings minimally charged under the two $\IZ_p$ factors produces $r$ F1s, with 
\beqa
r=\Delta\cdot\Delta'=\sum_k \frac{p_kp_l}{p^2}\, \Pi_k'\cdot \Pi_l'
\eeqa
The symmetry is a discrete Heisenberg group generated by elements $T$, $T'$ and a central element $C$, with relations
\beqa
T^p={T'}^{p'}=1\quad , \quad\quad TT'=C^r\, T'T
\label{heis1}
\eeqa
In principle, the element $C$ contains a finite order piece, since F1s carry discrete charges, as follows from a further 4d $BF$ coupling from the 10d Chern-Simons term 
\beqa
\int_{10d} B_2 \wedge {\ov F}_2\wedge F_6 \quad \to \quad \int_{4d} B_2 \sum_k  p_k \hat F_2^k
\eeqa
where $B_2$ is the NSNS 2-form, and
\beqa
&\int_{\Pi_k'\times a\times b} F_6=\hat F_2^k  
\eeqa
This kind of configuration has already appeared in section \ref{sec:IIAform2}, and leads to a $\IZ_q$ discrete symmetry, with $q=\frac{\sum_k (p_k)^2}{p}$.
This suggests the relation $C^q=1$, on top of (\ref{heis1}).

However, $C$ is actually slightly more subtle and involves the continuous part of the group. Indeed, the above group relations imply
\beqa
T'=T^pT'=C^{pr} T'T^p=C^{pr}T'
\label{multi-cross}
\eeqa
which, if $C$ involves just the discrete part of the symmetry, would require $pr=0$ mod $q$, which is not true in general. The point becomes clearer in the physical interpretation of (\ref{multi-cross}). Consider crossing one 4d string associated to $T'$ with $p$ 4d strings associated to $T$, leading to the creation of $pr$ F1s. Since the set of $p$ $T$-strings is trivial, we would expect the set of $pr$ F1s to be so. Physically, one can indeed annihilate the F1s in sets of $p$, by using combinations of D2-branes on $\Pi_k$ (each annihilating sets of $p_k$ strings).  However, as noted in section \ref{sec:IIAform2}, such D2-branes carry non-trivial monopole charge under the unbroken $U(1)$'s. Hence, the central element $C$ contains not only the discrete gauge transformation associated to the fundamental strings, but also a (dual) gauge transformation of the unbroken $U(1)$'s.

\section{Combining fluxes} 
\label{sec:combine}

When several kinds of fluxes are simultaneously present in a compactification, as often required by the equations of motion, inconsistent configurations of ${\bf Z}_p$-valued wrapped branes may naively arise. In a consistent microscopic theory like string theory such configurations should therefore not be possible. In this section we discuss some of these incompatibilities arising when combining several kinds of fluxes and the mechanisms by which string theory avoids such configurations in consistent flux compactifications.

\subsection{NSNS and RR fluxes: a puzzle and its resolution}

We first consider the combination of NSNS and RR fluxes. For concreteness, we focus on type IIB compactifications with simultaneous NSNS and RR 3-form fluxes, since this is a particularly popular setup for moduli stabilization; similar lessons apply to other type IIA or IIB setups with NSNS and RR fluxes. 

Consider the system of section \ref{sec:IIB}, for simplicity with only one 3-cycle $\alpha$ and its dual $\beta$, and introduce NSNS and RR 3-form fluxes such that
\begin{equation}
\int_{\alpha} {\ov F}_3=p \ , \qquad \int_{\beta} {\ov H}_3=p'
\end{equation}
Performing dimensional reduction, these fluxes lead to the 4d $BF$ couplings
\beqa
\int_{4d} \, \left(\, p\, B_2\wedge \hat F_2\, -\, p'\, C_2\wedge \hat F_2'\, \right)\label{bfhf}
\eeqa
where we have defined the 4d field strengths
\begin{equation}
\int_{\beta} F_5=\hat F_2 \ ,\qquad \int_{\alpha} F_5=\hat F_2'
\end{equation}
Combining results from section \ref{sec:IIB}, the first $BF$ coupling in eq.~(\ref{bfhf}) describes a $\IZ_p$ gauge symmetry with particles given by D3-branes on $\beta$ (annihilating on an NS5-brane on $\IX_6$) and strings from F1s (annihilating on a D3-brane on $\alpha$); while the second $BF$ coupling describes a $\IZ_{p'}$ symmetry with particles from D3-branes on $\alpha$ (annihilating on a D5-brane on $\IX_6$) and strings from D1-branes (annihilating on a D3-brane on $\beta$).

This naive combination however leads to some puzzles, as follows. Charged particles under the $\IZ_p$ symmetry play the role of string junctions under the $\IZ_{p'}$ symmetry, and vice versa\footnote{Microscopically, D3-branes on $\beta$ have actually a FW anomaly and emit D-strings with $\IZ_{p'}$ charge, and conversely, D3-branes on $\alpha$ emit fundamental strings with $\IZ_p$-charge.}. This naively allows configurations where $pp'$ $\IZ_{p'}$-charged strings annihilate on $p$ $\IZ_p$-charged particles, that annihilate on an instanton; or  conversely, $pp'$ $\IZ_{p}$-charged strings annihilating on $p'$ ${\IZ}_{p'}$-charged particles, that annihilate on an instanton. Such configurations must however be inconsistent, as the boundary of a string cannot have a boundary itself.

We can rephrase the problem by recalling from appendix \ref{sec:field-th} that string junctions are ``magnetic monopoles'' of the broken $U(1)$ gauge symmetries. The double role of $\IZ_p$ electrically charged particles playing as  $\IZ_{p'}$ string junctions reflects an underlying simultaneous gauging of a gauge boson and its magnetic dual. This also follows from the 4d electric-magnetic duality between $\hat F_2$ and $\hat F_2'$, that arises from the 10d self-duality of the RR 4-form. Such gaugings are inconsistent already on purely 4d grounds: denoting by $\phi$ the 4d scalar dual to the 2-form $B_2$, the $BF$ couplings (\ref{bfhf}) imply the gauge transformations
\beqa
&& (a) \;A_1\to A_1+d\lambda \quad ,\quad \phi\to \phi+p\lambda \nonumber \\
&& (b) \;C_2\to C_2+d\Sigma_1 \quad ,\quad A_1\to A_1+p'\,\Sigma_1 \label{incons}
\eeqa
However, it is not possible to write a consistent Lagrangian that respect both transformations and the relation $\hat F_2=*_4\hat F_2'$. In physical terms, (\ref{incons}a) describes the gauge boson $A_1$ becoming massive by eating up the scalar $\phi$, while (\ref{incons}b) describes the 2-form $C_2$ becoming massive by eating up the presumed massless gauge boson $A_1$, which is in fact massive and therefore contains too many degrees of freedom.\footnote{This problem has also been pointed out in the supergravity literature, in the context of the embedding tensor formalism (see e.g.~\cite{deWit:2005ub}). In that case, consistency of the theory requires the embedding tensor to satisfy a quadratic constraint which ensures that the gauging can be turned into a purely electric one in a suitable symplectic frame. In the above discussion we are dealing with genuinely electric/magnetic gaugings, which cannot be rotated into electric ones and which therefore lead to inconsistencies in the 4d theory.}

The above discussion does not imply that type IIB vacua with NSNS and RR fluxes are inconsistent, but rather that string theory must include extra ingredients to circumvent these problems. Indeed, it is familiar that the combination of NSNS and RR fluxes contributes to the RR tadpoles cancellation conditions. In the above type IIB case there is a tadpole of D3-brane charge given by the suggestive amount
\beqa
N_{\rm flux}=\int_{\IX_6} {\ov F}_3\wedge {\ov H}_3\, =\, pp'
\eeqa
It turns out that the extra ingredients required to cancel the tadpole precisely solve the above inconsistencies. We consider here three possibilities:

\medskip

{\bf - Orientifold planes.} The above tadpole can be cancelled by introducing O3-planes, as often done in the context of moduli stabilization. Their effect on the fields relevant to the discrete symmetries is drastic, since they are all projected out and no remnant discrete symmetry is left. The above problems are solved by removing degrees of freedom and rendering the structures trivial.

\medskip

{\bf - Anti-branes.} A second possibility is to introduce $\overline{\rm D3}$- (or D3-)branes. They modify the above discussion because their overall worldvolume $U(1)$ couples to the relevant 2-forms through the $\overline{\rm D3}$-brane CS couplings. Denoting respectively by $f_2$ and $f_2'$ the field strength of the overall $U(1)$ and its 4d dual on a stack of $pp'$ branes, we have the coupling
\beqa
-\int_{4d} pp'\, (\, C_2 \wedge f_2\, +\, B_2 \wedge f_2'\, )\label{anti}
\eeqa
The appearance of new degrees of freedom solves the problems, and leads to non-trivial discrete gauge symmetries as follows. Let us denote by $Q_{[\beta]}$ and $Q_{[\alpha]}$ the electric and magnetic generators that correspond to $\hat F_2$ and $\hat F_2'$ in eq.~(\ref{bfhf}), respectively; and $Q_{U(1)_e}$ and $Q_{U(1)_m}$ the electric and magnetic generators of the $\overline{\rm D3}$-brane $U(1)$, corresponding to the field strengths $f_2'$ and $f_2$ in eq.~(\ref{anti}). An NS5-brane instanton violates these $U(1)$ charges by $\Delta Q_{[\beta]}=p$ and $\Delta Q_{U(1)_e}=-pp'$, while a D5-brane instanton gives $\Delta Q_{[\alpha]}=-p'$ and $\Delta Q_{U(1)_m}=-pp'$. Now consider the linear combinations
\beqa
Q_1=p\,Q_{[\beta]}-Q_{U(1)_e}\quad ,\quad Q_2=p'\,Q_{[\alpha]}+Q_{U(1)_m}
\eeqa
The magnetic dual of $Q_1$ is $p\,Q_{[\alpha]}\!-\!Q_{U(1)_m}$,  which is preserved by all instantons. Hence, the monopoles of $Q_1$ do not decay, an can play the role of junctions for strings. These are associated to the discrete subgroup $\IZ_{p^2-pp'}\subset U(1)_{Q_1}$ preserved by the NS5-brane instantons. Similarly $U(1)_{Q_2}$ leads to a discrete $\IZ_{p'{}^2+pp'}$ symmetry.

\medskip

{\bf - Non-compact throats.} A third possibility is to consider non-compact setups, without the introduction of O3-planes or extra branes. This is considered in \cite{Copeland:2003bj} for warped throats \cite{Klebanov:2000hb}. In the above simple example this corresponds to taking the limit of large size for e.g. the 3-cycle $\beta$. There are surviving $\IZ_p$-charged strings, reflecting a local version of the $\IZ_p$ gauge symmetry, while there is no remnant of the $\IZ_{p'}$ symmetry, hence avoiding problems.\\

\subsection{Purely RR fluxes and symplectic rotations}

There are combinations of fluxes that do not lead to tadpole contributions. For instance, consider a type IIA compactification with $p'$ units of ${\ov F}_0$ and $p$ units of ${\ov F}_6$ flux. From section \ref{ch2}, there are $\IZ_{p'}$-charged particles arising from D6-branes on $\IX_6$ (annihilating on an NS5-brane on $\IX_6$) and $\IZ_{p'}$-charged strings from fundamental strings (annihilating on D0-branes). In addition, there are $\IZ_{p}$-charged particles from D0-branes (annihilating on an NS5-brane on $\IX_6$) and $\IZ_{p}$-charged strings from fundamental strings (annihilating on a D6-brane on $\IX_6)$. The system naively suffers from the troubles of the previous section, since $\IZ_{p'}$ electrically charged particles are also junctions for $\IZ_{p}$-charged strings, and vice versa. In the present case, however, the combination of ${\ov F}_0$ and ${\ov F}_6$ fluxes does not contribute to the RR tadpoles and no extra ingredient can come to the rescue.

Despite the above superficial analogy, this configuration differs from the previous subsection in that $\IZ_p$ and $\IZ_{p'}$-charged strings are both given by fundamental strings, and instantons are NS5-branes on $\IX_6$ in both cases as well. This in particular implies that there is only one scalar that is being gauged by the gauge potential. Concretely, the CS terms read
\beqa
\int_{10d} \big(\, {\ov F}_0 B_2\wedge F_8\, +\, {\ov F}_6 \wedge B_2\wedge F_2\,\big) \; \to\; 
\int_{4d} \big(\, p'\, B_2\wedge \hat{F}_2' + p\, B_2\wedge F_2\, \big)\, 
\eeqa
with $\hat{F}_2'=\int_{\IX_6}F_8$ the 4d dual of $F_2$. In contrast with the previous section, this describes a gauging by a combination of the electric and magnetic gauge potentials, that can be turned into a purely electric one by a 4d electric-magnetic symplectic transformation. Extracting $r={\rm gcd}(p,p')$ we have
\beqa
\int_{4d} \, r B_2\wedge \Big(\, \frac{p'}{r} \hat{F}_2' + \frac{p}{r} F_2\, \Big)\; \stackrel{\rm sympl.}{\longrightarrow} \; \int_{4d} \, r B_2\wedge f_2
\eeqa
where $f_2$ is the field strength associated to the combination $Q=(p'/r) Q_e+(p/r)Q_m$ of the electric and magnetic charge generators. There is a  $\IZ_r$ discrete gauge symmetry, nicely matching the annihilation/creation processes as follows. An NS5-brane instanton annihilates $p'$ D6- and $p$ D0-branes, namely $r$ sets of the basic unit of $Q$-charge ($\frac{p'}r$ D6- and $\frac{p}r$ D0-branes). Similarly, $\IZ_r$-charged strings are fundamental strings that can annihilate in sets of $p'$ on a D0-brane junction or in sets of $p$ on a wrapped D6-brane junction, and are hence conserved modulo $r$.

Similar conclusions apply to other combinations of type IIA RR fluxes. The above discussion is in fact straightforward in the mirror IIB setup, in which all RR fluxes map into 3-form flux. In particular, 4d electric-magnetic symplectic transformations are simply changes in the symplectic basis of 3-cycles $\alpha$ and $\beta$ in the IIB picture.

\section{Strings and unstable domain walls}
\label{sec:udw}

We have seen in the previous section that the orientifold in supersymmetric flux compactifications generically projects out all flux-induced $BF$ couplings and therefore also 4d strings ending on string junctions. There is however another set of discrete (${\bf Z}_p$-valued) brane wrappings in supersymmetric flux compactifications which usually do survive the orientifold projection, namely 4d strings with domain walls attached to them. Nucleation of such 4d strings render some of the domain walls in flux compactifications  unstable\footnote{The term ``unstable'' is used in this and following sections in a merely topological sense and does not imply any dynamical content.} (see figure \ref{fig:unstable-dw}) and vacua separated by the wall, which would naively seem to be different, are actually identical up to discrete identifications. In this section we describe these 4d objects from a microscopic point of view.  We begin our analysis with a simple illustrative example to make the main ideas manifest. Subsequently we discuss similar phenomena in more general flux compactifications.

\subsection{$F_2$ flux quantization in massive IIA}
\label{sec:f2quant}

\begin{figure}[!ht]
\begin{center}
\includegraphics[scale=.4]{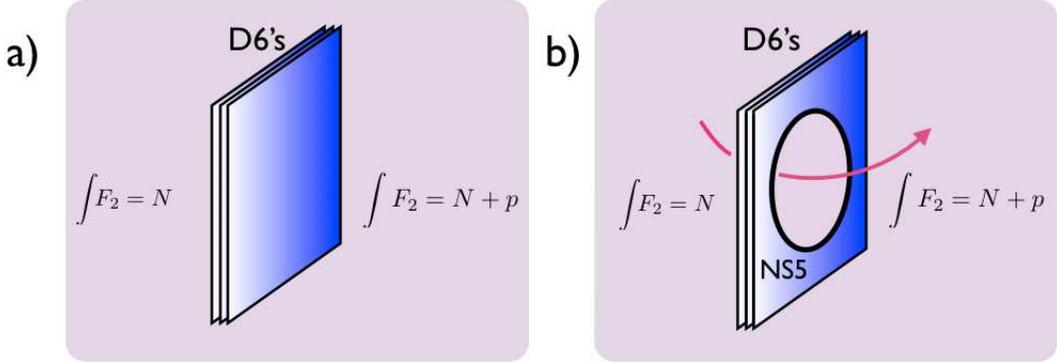}
\caption{\small a) A 4d domain wall obtained by $p$ D6-branes wrapped on a 4-cycle in a CY compactification of massive IIA theory. It separates two regions of 4d spacetime which differ by $p$ units of RR $F_2$ flux on the dual 2-cycle. b) The domain wall is unstable by nucleation of holes bounded by strings, realized as one NS5-brane wrapped on the 4-cycle. The two vacua with differing flux must therefore be equivalent.\label{fig:unstable-dw}}
\end{center}
\end{figure}

In order to introduce some of the main ideas, let us consider massive IIA theory compactified on a 6d manifold with mass parameter ${\ov F_0}=p$. Besides the $\IZ_p$-valued fundamental strings that we discussed in section \ref{sec:IIAform0}, the system admits also a set of 4d strings arising from NS5-branes wrapped on 4-cycles. The FW anomaly on the NS5-brane forces these strings to have domain walls attached to them, given by $p$ D6-branes wrapped on the same 4-cycle and ending on the NS5-brane. Hence, $p$ such D6-brane domain walls are unstable against nucleation of a string loop, as depicted in figure \ref{fig:unstable-dw}. 

From standard arguments  \cite{Taylor:1999ii}, two vacua separated by a single domain wall differ in one unit of RR flux ${\ov F}_2$ along the 2-cycle dual to the wrapped 4-cycle. The above instability of a set of $p$ domain walls therefore indicates that vacua differing by $p$ units of ${\ov F}_2$ flux along the 2-cycle are actually equivalent. Using the argument for different 2-cycles, this implies the quantization condition
\beqa
p_k\equiv \int_{\Pi_k} {\ov F}_2\, \in \, p\,\IZ\quad \mbox{for any $\Pi_k$}\in H_2(\IX_6,\IZ)
\eeqa
so the flux is $\IZ_p$-valued, rather than $\IZ$-valued. This deviation from the cohomological classification of RR fields should be a generalization (suitable for the presence of 0-form flux) of the K-theory classification of RR fields \cite{Moore:1999gb}. It would be interesting to derive this peculiar quantization from a holographic field theory dual in the context of the massive IIA AdS$_4$/CFT$_3$ duals uncovered in \cite{Gaiotto:2009mv}.

Unstable domain walls typically arise in theories on which a discrete gauge symmetry $G$ is spontaneously broken to a subgroup $H$ (see e.g.~\cite{Lee:1994qg} for a review). This description encompasses our example by considering the group $G={\bf Z}$ of monodromies generated by the 4d strings, which is broken by ${\ov F_0}$ to a subgroup $H={\bf Z}_p$. Strings with monodromies $a\in G/H$ which lay in the broken generators of $G$ cannot be stable and have attached domain walls. Those are therefore classified by the cosets $aH$, so that different strings can bound the same domain wall only if they belong to the same coset. In cases where $G$ contains several factors, these properties lead to an interesting interplay between the different types of 4d strings and domain walls, as we show in what follows for more general IIA flux vacua.

\subsection{$\IZ_p$-valued domain walls in IIA flux vacua}
\label{sec:zpdwiia}

In a fully consistent flux compactification the equations of motion typically demand different kinds of fluxes to be simultaneously switched on. Unstable domain walls, like those in the previous subsection, can produce intricate identifications of seemingly different flux backgrounds. From the point of view of the effective supergravity, however, these equivalences follow from simple underlying axion-like field identifications. Indeed, in the above example of D6-brane domain walls, the introduction of ${\ov F}_0=p$ in type IIA theory implies the presence of a Chern-Simons coupling in the definition of the physical RR 2-form  (see e.g. \cite{Polchinski:1998rr})
\begin{equation}
\tilde F_2 = dC_1+{\ov F}_2 + pB_2\label{g2}
\end{equation}
When crossing a domain wall through a hole bounded by a 4d string, as in figure \ref{fig:unstable-dw}b, the axion $\phi_k=\int_{\Pi_k} B_2$ experiences thus a monodromy $\phi_k\to \phi_k+1$ and, in addition a shift in the ${\ov F}_2$ flux, $p_k\to p_k-p$. According to eq.~(\ref{g2}), these two effects cancel each other in the sense that the physical field strength $\tilde F_2$ is left invariant, which implies that both vacua are equivalent.

This argument can easily be extended to arbitrary flux backgrounds. The complete set of physical field strengths that appear in the 10d type IIA supergravity action is
\begin{equation}
H_3 = dB_2+\bar H_3\ , \quad \tilde F_p=dC_{p-1}-H_3\wedge C_{p-3}+(\bar Fe^{B_2})_p\label{generalf}
\end{equation}
where $\bar F=\bar F_0+\bar F_2+\bar F_4 +\bar F_6$ and $(\cdot)_p$ selects the $p$-form component of a poly-form. In absence of 1-cycles, the axion-like fields of the compactification come from dimensionally reducing  $B_2$ along the 2-cycles and $C_3$ along the 3-cycles of the compact manifold. Strings in 4d are thus given by NS5-branes wrapping 4-cycles and D4-branes wrapping 3-cycles, as those induce the correct monodromy for the corresponding axions. In the presence of general flux backgrounds these objects develop FW anomalies, that are cancelled by attaching to domain walls. From eq.~(\ref{generalf}) it is straightforward to work out the axion-like identifications induced by the flux background and to interpret them in terms of unstable domain walls, as we did above for $\tilde F_2$. To linear order in the shifts, we obtain the structure of unstable domain walls summarized in table \ref{dwtable}.

\begin{table}[!ht]
\begin{center}
\begin{tabular}{|c|c||c|c||c|}
\hline
\multicolumn{2}{|c||}{\bf Domain wall}&\multicolumn{2}{|c||}{\bf String} & {\bf Rank}\\
\hline
type & cycle & type & cycle & \\
\hline\hline
D2 & $-$ & NS5 & $[\Gamma_4]\in H_4({\bf X}_6,\IZ)$ & $\int_{\Gamma_4}\overline{F}_4$\\
\hline
D4 & $[\Pi_2]\in H_2({\bf X}_6,\IZ)$ & NS5 & $[\Gamma_4]\in H_4({\bf X}_6,\IZ)$ & $\int_{\Gamma_4}\overline{F}_2\wedge \pi_2$\\
\hline
D6 & $[\Gamma_4']\in H_4({\bf X}_6,\IZ)$ & NS5 & $[\Gamma_4]\in H_4({\bf X}_6,\IZ)$ & $\int_{\Gamma_4}\overline{F}_0\pi_4$\\
\hline
D2 & $-$ & D4 & $[\alpha_3]\in H_3({\bf X}_6,\IZ)$ & $\int_{\alpha_3}\overline{H}_3$\\
\hline
\end{tabular}
\caption{Domain walls in flux compactifications and the type of 4d strings that can nucleate in the presence of fluxes. The last column denotes the number of domain walls that are needed to nucleate a hole bounded by a string.\label{dwtable}}
\end{center}
\end{table}

Having multiple types of domain walls and strings allows for new phenomena that were not present in the simple example of the previous subsection. In particular:\\

\begin{figure}[!ht]
\begin{center}
\includegraphics[scale=.4]{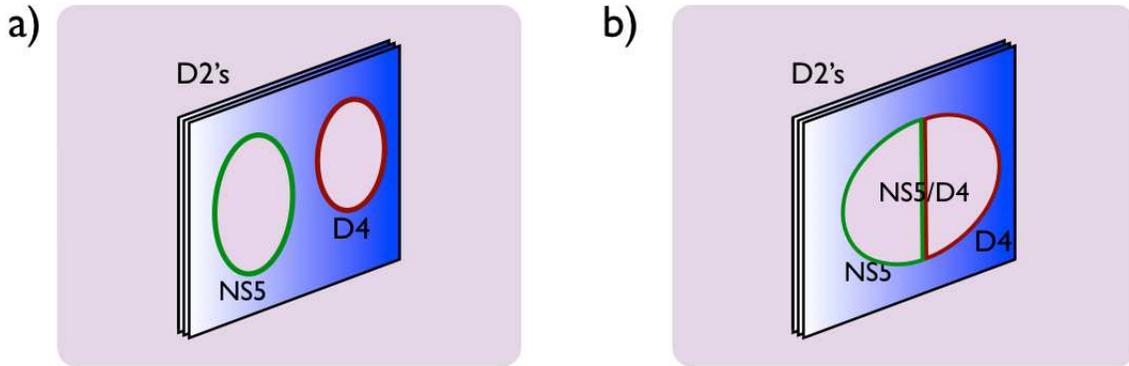}
\caption{\small Collision of holes in unstable D2-brane domain walls in 4d compactifications with ${\ov F}_4$ and ${\ov H}_3$ fluxes.\label{fig:holes}}
\end{center}
\end{figure}

{\bf - Hole collisions.} If a domain wall can decay via nucleation of different strings, there can be collisions of different types of holes \cite{Lee:1994qg}. This is the case of D2-brane domain walls in IIA compactifications with $\overline{F}_4$ and $\overline{H}_3$ fluxes, as they can decay via nucleation of 4d strings from NS5-branes wrapping a 4-cycle or from D4-branes wrapping a 3-cycle (see table \ref{dwtable}). The collision of the corresponding holes leads to a configuration with a single hole, crossed by a new 4d string, as shown in figure \ref{fig:holes}. Such string is a NS5/D4 bound state with cancelled FW anomalies and therefore has no domain wall attached. In terms of the M-theory uplift the new 4d string corresponds to a single M5-brane wrapping a linear combination of 4-cycles such that the total $G_4$ flux on it cancels.\\

\begin{figure}[!ht]
\begin{center}
\includegraphics[scale=.3]{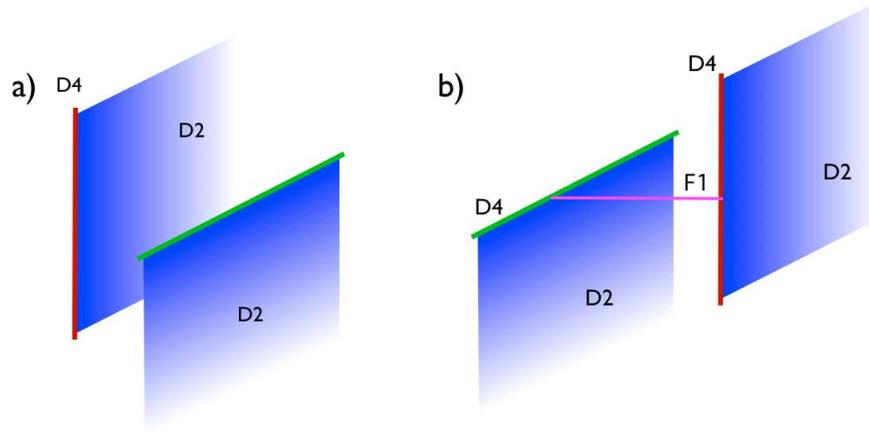}
\caption{\small In compactifications with ${\ov H}_3$ fluxes, crossing of 4d strings (wrapped D4-branes) with attached domain walls (D2-branes) produces 4d strings with no domain wall (F1s).\label{fig:dw-hw}}
\end{center}
\end{figure}

{\bf - Hanany-Witten effect for strings bounded with walls.} In theories where the unstable domain walls are associated to the spontaneous breaking of a non-Abelian discrete symmetry group $G$ (to a subgroup $H$) there is an interesting interplay between the 4d strings attached to domain walls of this section and the 4d strings that were discussed in section \ref{ch2}. Indeed, consider two 4d strings with non-commuting monodromies $a$ and $b$ associated to two broken generators of $G$, and therefore with attached domain walls. If the commutator $c=a^{-1}b^{-1}ab$ lies in $H$, then the crossing of the two strings leads to the creation of a new stretched 4d string with monodromy $c$ with no domain wall attached. This situation arises in type IIA flux compactifications without orientifold planes. To be more precise, consider a compactification with $\overline{H}_3$ flux on two hodge dual 3-cycles, $\alpha_3$ and $\beta_3$. From table \ref{dwtable} we see that a D4-brane wrapped on $\alpha_3$ leads to a 4d string with $\int_{\alpha_3}\overline{H}_3$ D2-brane domain walls attached, and a similar statement holds for a D4-brane wrapped on $\beta_3$. By the HW effect, crossing these 4d strings results in a stretched F1 with no domain wall attached, see figure \ref{fig:dw-hw}. 

\subsection{Type IIB $SL(2,\IZ)$ from unstable domain walls}
\label{sec:sdual-dw}

Unstable domain walls are also present in type IIB compactifications with NSNS and RR 3-form fluxes. We keep their discussion brief, since it is similar to the type IIA case. We consider a generic type IIB compactification on a Calabi-Yau ${\bf X}_6$ with NSNS and RR 3-form fluxes, and for simplicity, restrict to a single 3-cycle $\alpha_3$ and its hodge dual $\beta_3$ with general fluxes
\beqa
& \int_{\alpha_3}{\ov H}_3=N\quad & \quad \int_{\beta_3} {\ov H}_3=M\nonumber \\
& \int_{\alpha_3}{\ov F}_3=N'\quad & \quad \int_{\beta_3} {\ov F}_3=M'
\label{typeIIBex}
\eeqa
From by now familiar arguments, a D7-brane wrapping ${\bf X}_6$ leads to a 4d string with $k=\textrm{gcd}(N,M)$ domain walls attached. The latter consist of D5-branes wrapped on the class $(1/k)(M[\alpha_3]-N[\beta_3])$ and are required in order to cancel the ${\ov H}_3$-induced FW anomaly on the D7-branes. Hence, a set of $k$ such domain walls can decay by nucleation of a 4d string, as in figure \ref{fig:unstable-dw} (with the obvious replacements). As one crosses the domain wall, the complex axion-dilaton and the 3-form fluxes experience a $SL(2,{\bf Z})$ transformation
\beqa
\tau \to \tau + 1 \ , \quad N'\to N'+N\ , \quad M'\to M'+M
\label{flux-action}
\eeqa
This can be extended to the entire type IIB $SL(2,{\bf Z})$ duality group by considering (sets of) 4d strings arising from $(p,q)$ 7-branes wrapped on ${\bf X}_6$ and domain walls arising from general bound states of D5 and NS5-branes wrapped on 3-cycles. Hence, unstable domain walls encode the existence of  non-trivial dualities in string theory. 

Crossing these unstable domain walls leaves always invariant the contribution of the 3-form fluxes to the RR 4-form tadpole
\beqa
N_{\rm flux}\equiv \int_{\IX_6} {\ov H}_3\wedge {\ov F}_3=NM'-N'M
\label{nflux}
\eeqa
and therefore the number of O3-planes, D3-branes or other sources required for tadpole cancellation. This is however not true in general for the more familiar case of \emph{stable} domain walls, which change the flux background in a way that cannot be undone by a  $SL(2,{\bf Z})$ transformation\footnote{See e.g.  \cite{Kachru:2002ns} for an explicit discussion.}. For instance, a domain wall given by one D5-brane on $\alpha_3$ shifts the ${\ov F}_3$ flux on $\beta_3$ by one unit. The change in the 3-form flux contribution to the RR tadpoles is compensated by the appearance of $N$ spacetime filling D3-branes on one side of the domain wall, which are microscopically required by a FW anomaly on the D5-brane domain wall.

\subsection{$\IZ_p$-valued domain walls and axion stabilization}

From a four-dimensional viewpoint, one can associate the presence of $\IZ_p$-valued domain walls with the stabilisation of axions in the effective theory. In brief, the same flux that renders the domain walls unstable also induces an effective potential for the axion associated to the unstabilizing string, and the different axionic vacua are connected by $\IZ_p$-valued domain walls.

To illustrate this point let us again consider the last example of type IIB compactifications with 3-form fluxes. It is well-known that the fluxes generate an effective potential for the closed string light modes, encoded in the superpotential \cite{Gukov:1999ya}
\beq
W_{\rm GVW}\, =\, \int_{\IX_6} \Omega \wedge ({\ov F}_3 - \tau {\ov H}_3)\, =\, \int_{\IX_6} \Omega \wedge (\tilde{F}_3 - i g_s^{-1} {\ov H}_3)
\label{supoGVW}
\eeq
with $\tilde{F}_3 = {\ov F}_3 - C_0 {\ov H}_3$. In the language of section \ref{sec:zpdwiia}, the transformation (\ref{flux-action}) is 
\beq
C_0 \to C_0 + 1 \ , \quad {\ov F}_3 \to {\ov F}_3 + {\ov H}_3
\label{shiftIIb}
\eeq
which not only leaves invariant (\ref{nflux}), but also the physical field strength $\tilde{F}_3$ and so the superpotential (\ref{supoGVW}). Notice that $W_{\rm GVW}$ is left invariant off-shell, so this $SL(2,{\bf Z})$ identification is indeed independent of supersymmetry. 

In this example (\ref{typeIIBex}), considering $k=\textrm{gcd}(N,M)>1$ amounts to say that ${\ov H}_3$ is $k$ times the integer 3-form $h_3 = (N/k) \a_3 + (M/k) \b_3$, or in general that ${\ov H}_3 = k\, h_3$, with $h_3 \in H^3(\IX_6, {\bf Z})$. It is natural to generalize (\ref{shiftIIb}) to include fractional shifts
\beq
C_0 \to C_0 + \frac{1}{k} \ , \quad {\ov F}_3 \to {\ov F}_3 +  \frac{1}{k} {\ov H}_3
\label{shiftIIbb}
\eeq
which again leave invariant $\tilde{F}_3$, $W_{\rm GVW}$ and $N_{\rm flux}$. These fractional axionic shifts do not correspond to equivalent type IIB flux vacua, since the axion $C_0$ takes inequivalent values before and after the shift. They can instead be associated with $\IZ_k$-valued domain walls, which interpolate between these $k$ vacua with different values for the axion $C_0$. Indeed, let us assume that we are at the minimum of the scalar potential generated by the fluxes (\ref{typeIIBex}). A D5-brane wrapped on the homology class $(1/k)(M[\alpha_3]-N[\beta_3])$ will shift the value of the flux ${\ov F}_3$ by the amount $h_3 = (N/k) \a_3 + (M/k) \b_3$ and take the system away from the potential minimum. In order to attain a new vacuum, we only need to shift the axion $C_0$ by the amount of $1/k$, as the combined effect (\ref{shiftIIb}) takes us back to the previous value of the background flux $\tilde{F}_3$, that the superpotential and scalar potential depend on. 

In general, in type IIB compactifications with 3-form fluxes one can directly relate the fact that the RR axion $C_0$ is stabilized with the presence of $\IZ_k$-valued domain walls. Such domain walls will connect vacua whose background is identical except for the value of $C_0$, which is a multiple of $1/k$. The same kind of statement applies to the $\IZ_p$-valued domain walls that have been discussed in the context of type IIA flux compactifications, the main difference being that now we have more than one stabilized axion and so more than one class of unstable domain wall. In this case the superpotential generated by RR fluxes is given by \cite{Gukov:1999gr}
\beq
W_{\rm IIA, RR}\, =\, \int_{\IX_6} e^{iJ} \wedge \left({\ov F}_0 + \tilde{F}_2 + \tilde{F}_4 + \tilde{F}_6 \right)
\label{IIAsupo}
\eeq
and the axions stabilized by it are $\phi_k=\int_{\Pi_{2,k}} B_2$ and $\varphi_j=\int_{\alpha_{3,j}} C_3$. The set of $\IZ_p$-valued domain walls that correspond to these axions are the ones in table \ref{dwtable}, and they interpolate between the finite set of vacua that have the same values for $\tilde{F}_{p}$ but differ by a fractional value for the axions $\phi_k$ and $\varphi_j$.

In the flux literature, the IIA superpotential (\ref{IIAsupo}) is usually analyzed in the context of compactifications that also include geometric fluxes. While so far we have not considered the presence of the latter, one can show that they also imply the presence of discrete branes that then show up as $\IZ_p$-valued particles, strings and domain walls in 4d. Hence, geometric fluxes do also fit into the general discussion of discrete wrapped branes, as we now turn to discuss.

\section{Geometric fluxes and torsion (co)homology}
\label{sec:torsion-hom}

As discussed in \cite{Camara:2011jg} (see also \cite{Gukov:1998kn} and \cite{BerasaluceGonzalez:2012vb}), compactification of $p$-form fields on spaces with torsion (co)homology classes leads to discrete gauge symmetries. In the context of flux compactifications, torsion arises in models with geometric fluxes \cite{Kachru:2002sk}, which are dual to field strength fluxes. In this section we describe several dualities between the flux configurations in earlier sections, and setups with geometric fluxes with discrete gauge symmetries arising from torsion (co)homology. Richer combinations of coexisting torsion and field strength fluxes are left for future work.

\subsection{M-theory lift of type IIA RR $F_2$-flux}

Consider the M-theory lift of type IIA models with RR ${\ov F}_2$ flux, see section \ref{sec:IIAform2}. We have a compactification on $\IX_6\times \IS^1$ with a geometric flux $\omega_2={\ov F}_2$ twisting the $\IS^1$ over 2-cycles in $\IX_ 6$. In other words, this is M-theory on a 7-manifold $\IX_7$ given by  an $\IS^1$ bundle over $\IX_6$ (note we have introduced neither O6-planes nor D6-branes) with first Chern class $c_1=\omega_2$ with
\beqa
\int_{\Pi_k} \omega_2\, =\, p_k
\eeqa
see eq (\ref{f2-int}). The twisting of the $\IS^1$ introduces torsion 1-cycles, and dual torsion 5-cycles, concretely $H_1(\IX_7,\IZ)=H_5(\IX_7,\IZ)=\IZ_q$ (with $q$ as in section \ref{sec:IIAform2}, although we will not enter these details). The basic torsion 1-cycle is the $\IS^1$ itself (since 2-cycles on the base $\IX_6$ are 2-chains in $\IX_7$, with boundary given by a multiple of the fiber class), and the 5-cycles arise from fibering the $\IS^1$ over 4-cycles in $\IX_6$ (with the base $\IX_6$ itself providing the corresponding 6-chain in $\IX_7$). There are charged particles, arising from M5-branes on the torsion 5-cycles
(descending to type IIA D4-branes on 4-cycles of $\IX_6$, annihilating on an NS5 wrapped on $\IX_6$), and charged strings from M2-branes on the torsion 1-cycles (descending to type IIA F1s, annihilating on D2-branes on 2-cycles of $\IX_6$), precisely as in section \ref{sec:IIAform2}.

\subsection{T-dual of NSNS 3-form flux}

Let us now consider type II models with NSNS 3-form flux ${\ov H}_3$, e.g. section \ref{sec:IIAform3}, in a compactification with a (isometric) circle direction $x$ on which we T-dualize. Decomposing ${\ov H}_3=\omega_2\, dx+\ldots$, the T-dual has geometric flux $\omega_2$, twisting the T-dual $\IS^1$. The T-dual geometry has torsion 1-cycles (the basic being the $\IS^1$ itself) and 4-cycles (from the $\IS^1$ times a 3-cycle), namely $H_1(\IX_6,\IZ)=H_4(\IX_6,\IZ)=\IZ_p$. The T-dual objects to those in section \ref{sec:IIAform3} are as follows. Charged particles are D1-branes wrapped on the torsion 1-cycles (annihilating on D1-branes on the corresponding 2-chain); the charged strings arise from NS5-branes on torsion 4-cycles (annihilating on NS5-branes on the 5-chain).

In the type IIB picture, there is an additional $\IZ_p$ discrete gauge symmetry with charged particles from F1s on the torsion 1-cycles, and charged strings from D5-branes on the torsion 4-cycles. In the original type IIA picture, charged particles are KK modes along the $\IS^1$ direction (and strings are D4-branes on 3-cycles). Namely, the underlying $U(1)$ gauge symmetry arises from an isometry generator of the compactification space, a possibility we had not included in section \ref{sec:IIAform3}.\footnote{See appendix \ref{sec:kk} for discrete symmetries arising from isometry via flux catalysis.}

As a final comment, we note that the appearance of discrete gauge symmetries from torsion homology and from flux catalysis can be combined together, even into a non-Abelian discrete group. A simple example can be obtained from the configuration in appendix \ref{sec:isom-nonab}, by a T-duality turning the NSNS 3-form flux into a geometric flux.

\subsection{Domain walls and superpotentials from torsion}

Torsion in (co)homology arises quite naturally in the context of type IIA flux compactifications to Minkowski, in the sense that the compactification manifold $\IX_6$ is not Calabi-Yau. More precisely one finds that $\IX_6$ is an SU(3)-structure manifold with a 3-form $\Omega$ such that dRe $\Omega \neq 0$  \cite{Grana:2005jc}. This has a number of consequences, like the fact that a D-brane can wrap a torsional 3-cycle $\pi_3$ with $[\pi_3] \in {\rm Tor\, } H_3 (\IX_6, \IZ)$ and still be BPS \cite{Marchesano:2006ns}, something forbidden in a Calabi-Yau. By analysing the spectrum of BPS D-branes one finds that such BPS torsion D-branes are typically present, and so for this class of compactifications it is quite general to find that, e.g. ${\rm Tor\, } H_3 (\IX_6, \IZ) = \IZ_k$. As we will now see, this implies the presence of $\IZ_k$-valued 4d domain walls, with similar phenomena to those discussed in the previous subsection. 

The effect of a non-trivial ${\rm Tor\, } H_3 (\IX_6, \IZ) = \IZ_k$ on 4d domain walls can be detected by means of the superpotential 
\beq
W_{\rm IIA, NS}\, = \, \int_{\IX_6}{\rm Re\, } \Omega \wedge {H}_3 \, =\, \int_{\IX_6}{\rm Re\, } \Omega \wedge \bar{H}_3 - \int_{\IX_6} {\rm Re\, } d\Omega \wedge B_2
\label{IIAsupo2}
\eeq
that needs to be considered together with (\ref{IIAsupo}) in the analysis of type IIA flux vacua. In particular, the second piece of the rhs of (\ref{IIAsupo2}) can stabilise 4d axions that arise from the dimensional reduction of $B_2$. Notice however that since $\Omega$ is globally well-defined Re\,$d\Omega$ is trivial in cohomology, and so $B_2$ needs to be a non-closed 2-form for this piece of the superpotential to be non-vanishing. Expanding $B_2$ in non-closed 2-forms is rather unusual when performing dimensional reduction in Calabi-Yau compactifications, although in compactifications where ${\rm Tor\, } H_3 (\IX_6, \IZ) = {\rm Tor\, } H_2 (\IX_6, \IZ) \neq 0$ this is necessary to include the physics of massive modes related to torsion (co)homology \cite{Camara:2011jg}. Such prescription also arises in the context of type IIA flux compactifications in twisted tori \cite{Camara:2005dc}, where 2-forms $\omega_2$ satisfying
\beq
d\omega_2 \, =\, k \beta_3
\label{torsionpform}
\eeq
need to be considered in a consistent truncation of the theory. If we now expand $B_2 = \phi\, \omega_2 + \dots$, then $\phi$ represents a 4d axion that enters into the superpotential (\ref{IIAsupo2}). 

The 3-form $\b_3$ in (\ref{torsionpform}) is a torsion integer 3-form, more precisely we have that $[\beta_3] \in {\rm Tor\, } H^3 (\IX_6, \IZ) = \IZ_k$ generates such torsion group. As such, we can switch on the background flux $\ov{H}_3$ along $\beta_3$, and this will also contribute to the superpotential (\ref{IIAsupo2}). In particular, we have that a shift of the form
\beq
{\phi} \to {\phi} + \frac{1}{k} \ , \quad \ov{H}_3 \to \ov{H}_3 +  \b_3
\label{shiftIIa}
\eeq
leaves ${H}_3$ and hence the superpotential (\ref{IIAsupo2}) invariant. This shift is similar to the fractional axionic shift (\ref{shiftIIbb}) in the sense that it connects $k$ different flux vacua where ${H}_3$ remains the same but ${\phi}$ takes different values. Consequently, there must exist $\IZ_k$-valued domain walls interpolating among such vacua; they indeed arise from NS5-branes wrapping a torsion 3-cycle $\pi_3$, such that $[\pi_3] \in {\rm Tor\, } H_3 (\IX_6, \IZ) = \IZ_k$ is the Poincar\'e dual of $[\b_3]$. By the definition of torsion homology, $k$ copies of $\pi_3$ can be connected by a 4-chain $\Sigma_4$. Hence, $k$ NS5-branes on $\pi_3$ can be connected in $\IX_6$ by an NS5-brane wrapped on $\Sigma_4$, which corresponds to a 4d string with $k$ attached  domain walls, similarly to figure \ref{fig:unstable-dw}.

\section{Final remarks}
\label{sec:conclu}

In this paper we have studied wrapped branes in flux compactifications,  in which topological flux-induced effects render the physical charges $\IZ_p$-valued, in contrast with the naive $\IZ$-valued (co)homological charge. The basic mechanism is a generalization of \cite{Maldacena:2001xj}, including NSNS and RR fluxes, and general NS- and D-branes. The resulting groups hence generalize twisted K-theory, and their physical construction provides an interesting avenue to explore the mathematical formulation of such groups.

The microscopic mechanism underlying the $\IZ_p$ charge is the Freed-Witten anomaly, in its dual avatars. Our analysis adds on a macroscopic interpretation for the $\IZ_p$'s using 4d effective field theory intuitions. For $\IZ_p$-charged particles and strings, we have displayed an underlying  $\IZ_p$ discrete gauge symmetry, manifest in 4d $BF$ couplings arising from the 10d Chern-Simons couplings. Particular examples of these effects have appeared in the AdS/CFT context; in addition, we have shown that the flux catalysis process can produce discrete gauge symmetries in F-theory. Finally, we have applied our intuitions to the construction of systems where the discrete group classifying branes charges is non-Abelian. These results continue the programme initiated in \cite{Camara:2011jg,BerasaluceGonzalez:2011wy,BerasaluceGonzalez:2012vb} to understand the sources of discrete gauge symmetries in string theory. 

We have also discussed $\IZ_p$-charged domain walls, and argued that they encode non-trivial duality properties of string theory, at the level of the 4d effective action. Specifically, they imply the equivalence (under integer shifts of suitable axions) of seemingly different flux vacua. Moreover, the analysis of the relevant axion directions suggests an interesting structure of vacua, which are inequivalent but related by fractional axion shifts. It would be interesting to apply these intuitions to question in the phenomenology of string theory axions and their stabilization.

\bigskip \bigskip 

\centerline{\bf \large Acknowledgments}

\medskip

We thank M. Goodsell, L. Ib\'a\~nez and T. Weigand for discussions.  
This work has been partially supported by the grants FPA2009-09017, FPA2009-07908, FPA2010-20807 and Consolider-CPAN (CSD2007-00042)  from the Spanish Ministry of Economy and Competitiveness, HEPHACOS-S2009/ESP1473 from the C.A. de Madrid, AGAUR 2009-SGR-168 from the Generalitat de Catalunya and the contract ``UNILHC" PITN-GA-2009-237920 of the European Commission. M.B-G. acknowledges the finantial support of the FPU grant AP2009-0327. F.M. is supported by the Ram\'on y Cajal programme through the grant RYC-2009-05096 and by the People Programme of FP7 (Marie Curie Auction) through the REA grant agreement PCIG10-GA-2011-304023.

\appendix

\section{Field theory description of Abelian discrete gauge symmetries}
\label{sec:field-th}

In this article we often look for particles and strings charged with respect to a $\IZ_p$ discrete symmetry. Following  \cite{Banks:2010zn} (see also 
\cite{Hellerman:2010fv} for an alternative viewpoint on discrete gauge symmetries), the symmetry is usually identified from the presence of a 4d coupling $p\, \hat B_2\wedge \hat F_2$, between the $U(1)$ field strength $\hat F_2$ and a 2-form  $\hat B_2$ (in the normalization that $\hat B_2$ is dual to a scalar with periodicity $1$, and that the minimal $U(1)$ charge is 1). 
The Lagrangian for a $\IZ_p$ gauge  theory can be described in terms of a scalar $\phi$ (4d dual to $\hat B_2$) and a 1-form $A_1$, with gauge invariance
\beqa
A_1\to A_1+d\lambda \quad ,\quad \phi\to \phi + p\,\lambda
\label{gaugeinv1}
\eeqa
or in terms of the 2-form $\hat B_2$ and the dual 1-form $V_1$, with gauge invariance
\beqa
\hat B_2\to \hat B_2+d\Lambda_1\quad ,\quad V_1\to V_1+p\,\Lambda_1
\label{gaugeinv2}
\eeqa
To facilitate the identification of particles and strings that are charged under the discrete symmetry, and objects annihilating them, we recall that \cite{Banks:2010zn}:

\medskip

{\bf -} $\IZ_p$-charged particles arise from objects electrically charged under $A_1$. They are described by a line operator $e^{i\int_L A_1}$, which creates a charged particle with worldline $L$.

\medskip

{\bf -} The instanton operator annihilating sets of $p$ charged particles is the object whose action contains a term linear in $\phi$. The corresponding operator is
\beqa
e^{-i\phi} e^{i\,p\int_LA_1}
\label{instanton}
\eeqa
where here $L$ is a curve starting at the point $P$ where the insertion of $e^{i\phi}$ occurs. The structure of (\ref{instanton}) is required by the gauge invariance (\ref{gaugeinv1}).

\medskip

{\bf -} $\IZ_p$-charged strings arise from objects electrically charged under $\hat B_2$. They are described by a surface operator $e^{i\int_C \hat B_2}$, creating a charged string with worldsheet $C$.

\medskip

{\bf -} The junction annihilating sets of $p$ charged strings is the object coupling electrically to $V_1$, namely the ``would-be" monopole of the original $U(1)$ theory (which, since the $U(1)$ is Higgsed, are actually confined, with the $\IZ_p$-charged strings playing as flux strings). The corresponding operator is
\beqa
e^{-i\int_L V_1}\, e^{i\,p\int_C \hat B_2}
\label{line-op}
\eeqa
where here $C$ is a surface with boundary $L$. The structure of (\ref{line-op}) is required by the gauge invariance (\ref{gaugeinv2}).

A question that arises in the string theory setup is whether any non-trivial flux remains after the decay of $p$ charged objects described by operators (\ref{instanton}), (\ref{line-op}). Actually, these operators  are trivial in the field theory, in the sense that they do not have long range effects \cite{Banks:2010zn}. Indeed, the equations of motion of the $BF$ theory force all integrals of $F_2$ and $H_3$ to vanish. Since the integral of e.g.~$H_3$ over a 3-cycle measures the amount of instantons inside it, this simply states that `naked' instantons are not allowed in the theory, as was clear from gauge invariance. In fact,  the actual dressed instantons do carry a non-trivial charge, measured by the physical field strength $G_3=dB_2+p\,d*\!F_2$; but it is accounted not by $H_3$, which vanishes, but by the electric current $*F_2$ of the $p$ particles ending on it. Hence the decay of the $p$ charged particles is to the vacuum, rather than to a putative different flux sector.

\section{Freed-Witten and Hanany-Witten effects}
\label{fw-hw}

In this appendix we collect the main brane topological effects used in the text.\\ 

{\bf Freed-Witten effects}

\medskip

{\bf - FW1.} A D$p$-brane with worldvolume $S_{p+1}$ with homologically non-trivial ${\ov H}_3|_S$ must emit D$(p-2)$-branes on the Poincar\'e dual class \cite{Maldacena:2001xj} (morally, along directions of $S_{p+1}$  transverse to ${\ov H}_3|_S$).

\medskip 
 
{\bf - FW2.} An NS5-brane with worldvolume $S_6$ with homologically non-trivial ${\ov F}_p|_S$ emits D$(6-p)$-branes spanning the Poincar\'e dual class. This follows  by considering FW1 for D5-branes with ${\ov H}_3$ emitting D3-branes, S-dualizing to NS5-branes with ${\ov F}_3$ emitting D3-branes, and T-dualizing to general ${\ov F}_p$. 

\medskip

{\bf - FW3.} A D$p$-brane with worldvolume $S$ with homologically non-trivial ${\ov F}_p$ emits F1s along the Poincar\'e dual class \cite{Witten:1998xy}. This follows by considering FW1 for D3-branes with  ${\ov H}_3$ emitting D1-branes, S-dualizing to D3-branes with ${\ov F}_3$ emitting F1s, and T-dualizing to general D$p$-branes with ${\ov F}_p$.\\

{\bf Hanany-Witten or brane creation effects}

\medskip

\begin{figure}[!ht]
\begin{center}
\includegraphics[scale=.4]{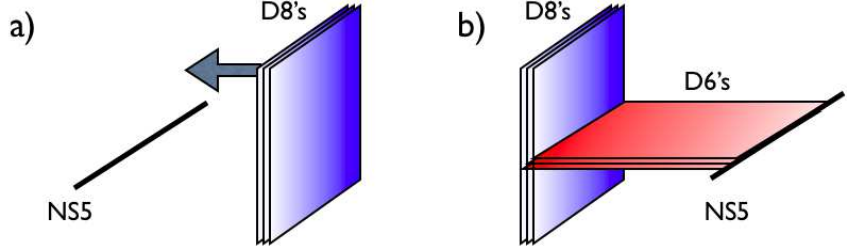}
\caption{\small Brane creation effect by crossing D8-branes and NS5-branes. Regarding the D8-brane as a source of ${\ov F}_0$ flux, it also shows that in the presence of a type IIA mass parameter ${\ov F}_0=p$, an NS5-brane can exist only with $p$ D6-branes ending on it.}
\label{fig:hwd6d8}
\end{center}
\end{figure}

{\bf - HW1.} An NS5-brane along directions $x^0, x^1,x^2,x^3,x^4,x^5$ and a D$(p+3)$-brane along directions $x^0,x^1, \ldots, x^p, x^6, x^7,x^8$, with $p\leq 5$ can be crossed in the direction $x^9$ leading to the creation of a D$(p+1)$-brane along $x^0,x^1,\ldots,x^p, x^9$ (see figure \ref{fig:hwd6d8} for the $p=5$ case). This effect follows from \cite{Hanany:1996ie} by T-duality (and coordinate relabelling). 

\medskip

\begin{figure}[!ht]
\begin{center}
\includegraphics[scale=.4]{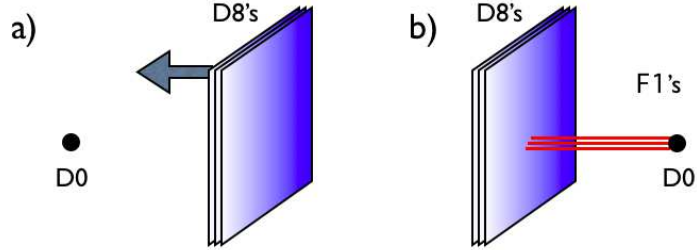}
\caption{\small Brane creation effect by crossing D8- and D0-branes. Regarding the D8-brane as a source of ${\ov F}_0$ flux, it also shows that in the presence of a type IIA mass parameter $p$, a D0-brane can exist only with $p$ fundamental strings ending on it.}
\label{fig:hwd0d8}
\end{center}
\end{figure}

{\bf - HW2.} A D$p$-brane along $x^0,\ldots, x^p$ and a D$(8-p)$-brane along $x^0, x^{p+1},\ldots, x^8$ can be crossed in $x^9$ leading to the creation of F1s in the directions $x^0,x^9$ (see  figure \ref{fig:hwd0d8} for the $p=0$ and $p=8$ cases). This follows by considering HW1 for creation of D1-branes from NS5- and D3-brane crossing, S-dualizing to the creation of F1s from D5- and D3-brane crossing, and T-duality to general $p$.

We conclude by mentioning that the FW and HW effects are related, as follows (see figures \ref{fig:hwd6d8}, \ref{fig:hwd0d8}). In the HW effect, we regard one of the branes as a source of flux, which the second brane (compactified by closing it at infinity) picks up in the form of FW anomaly, or not, depending on whether we are before or after the HW crossing.  The HW brane creation is required to cancel this FW anomaly, conversely the FW consistency condition is required to explain the HW brane creation.

\section{Flux catalysis and continuous isometries}
\label{sec:isometries}

In the main text we have focused on 4d $U(1)$ gauge bosons arising from higher dimensional $p$-form gauge potentials. In this appendix we briefly consider gaugings by KK gauge bosons from continuous $U(1)$ isometries in the compactification space. These do not arise in CY threefold compactifications, but may be present in other spaces. 

\subsection{A gauging by KK gauge bosons}
\label{sec:kk}

Consider e.g.~type IIB string theory compactified on a space $\IX_6$, which for simplicity we take $\IB_4\times \IT^2$. Let $x^4,\ldots, x^7$ denote (local) coordinates on $\IB_4$, and $x^8,x^9$ coordinates on $\IT^2$ (normalized with periodicity 1). We introduce $p$ units NSNS 3-form flux with one leg along $x^9$ and two legs on a 2-cycle $\Sigma_2$ on $\IB_4$ (locally along e.g. $x^6, x^7$),
\beqa
\int_{\Sigma_2\times (\IS^1)_9} {\ov H}_3=p
\eeqa
This flux breaks the KK $U(1)$ associated to $x^9$ because the 2-form gauge potential for such $H_3$ is not translationally invariant, as it can be written 
\beqa
B_2=p \,x^9\, dx^6 dx^7
\label{b2-gauge}
\eeqa
Upon dimensional reduction, there is a gauging of the scalar $\phi=\int_{\Sigma_2} B_2$, since a translation in $x^9$ shifts the value of $\phi$. Namely
\beqa
A_1\to A_1+d\lambda \quad ,\quad \phi\to \phi+p\lambda
\eeqa
This gauging defines a discrete $\IZ_p$ gauge symmetry from a discrete isometry (analogous to the gauging in magnetized D-branes in \cite{BerasaluceGonzalez:2012vb}). The gauging should be manifest as a $BF$ coupling upon dimensional reduction, although we will not need this result.

The particles that are charged under this discrete symmetry are states with KK momentum along $x^9$, and can annihilate in sets of $p$ on a fundamental string wrapped on $\Sigma_2$ (and localized in $x^9$, hence violating momentum conservation). The 4d $\IZ_p$-charged strings are NS5-branes on the 4-cycle dual to $\Sigma_2$ in $\IX_6$, namely along $x^4, x^5, x^8, x^9$. The junction annihilating $p$ strings is the KK monopole associated to $x^9$, namely a Taub-NUT (TN) geometry with isometry direction $x^9$, and base $\IR^3$ in the non-compact 4d space, see first three rows in table \ref{table:nonab}. Microscopically, the TN geometry can be shown to emit $p$ NS5-branes as follows. In the absence of the TN, ${\ov H}_3=p\, dx^6dx^7dx^9$, but when the TN is present $dx^9$ is not well-defined and must be promoted to $\rho_1\equiv dx^9+\vec{v}\cdot d\vec{x}$, where $\vec{x}=(x^1,x^2,x^3)$, and $\vec{v}$ is the Dirac monopole potential. This 1-form is not closed, but rather $d\rho_1=\omega_2$, where $\omega_2$ is a harmonic 2-form supported on the TN center. The Bianchi identity becomes
\beqa
dH_3\,=\,p\,\omega_2 \,dx^6dx^7
\eeqa
This source term induces a FW-like inconsistency, which must be cancelled by extra sources for $H_3$. These are $p$ NS5-branes along e.g. $x^0,x^1,x^4,x^5,x^8,x^9$, ending on the TN location in $x^1$ at a four-dimensional boundary along $x^0,x^4,x^5,x^8$ (notice that the $x^9$ direction shrinks at the TN center). 

The above argument is dual to a standard FW argument, as follows. T-duality along $x^8$ gives a type IIA configuration consisting of a TN geometry along $x^0,x^4,\ldots, x^8$ (with $\IS^1$ fiber along $x^{9}$), with ${\ov H}_3\sim p\, dx^6dx^7dx^9$ and NS5-branes along $x^0,x^1,x^4,x^5,x^8,x^9$. We can now perform a lift to M-theory by introducing a new direction denoted by $x^{9'}$, and shrink $x^9$ to get back to type IIA (a 9-9' flip). After this process, we end up with a D6-brane along $x^0,x^4,\ldots, x^8, x^{9'}$, with NSNS 3-form flux ${\ov H}_3\sim p\, dx^6dx^7dx^{9'}$ inducing a FW anomaly, cancelled by D4-branes along $x^0,x^1,x^4,x^5,x^8$. 
  
 \begin{table}[!ht]
\begin{center}
\begin{tabular}{c|cccc|cccccc}
\hline
NS5 & 0 & 1 & $\times$ & $\times$ & 4 & 5 & $\times$ &  $\times$ &  8 & 9\cr
$H_3$ & $\times$ & $\times$ & $\times$ & $\times$ & $\times$ & $\times$ & 6 & 7 & $\times$ & 9\cr
TN & 0 & $\times$ & $\times$ & $\times$ & 4 & 5 & 6 & 7 & 8 & $\otimes$ \cr
\hline
D5 & 0 & $\times$ & 2  & $\times$   &  4 & 5 & 6 & 7 & $\times$ & $\times$ \cr
$F_3$ & $\times$ & $\times$ & $\times$ & $\times$ & $\times$ & $\times$ & $\times$ & 7 & 8 & 9  \cr
TN & 0 & $\times$ & $\times$ & $\times$ & 4 & 5 & 6 & $\otimes$ & 8 & 9 \cr
\hline
\hline
D3 & 0 & $\times$ & $\times$ & 3 & 4 & 5 & $\times$ & $\times$ & $\times$ & $\times$ \cr
\hline
\end{tabular}
\end{center}
\caption{\small Relative geometry of the objects and fluxes involved in the realization of a non-Abelian discrete gauge symmetry. A number (resp. a cross) denotes that the corresponding brane or flux extends (or does not extend) along the corresponding direction. The circled cross indicates the isometric direction for Taub-NUT geometries producing line operators for string decays. In the top two triplets, the first row is the object producing the $\IZ_p$-charged string, while the second and third to the flux and junction catalyzing its decay. The last line describes the D3-branes corresponding to the 4d strings created upon crossing the NS5- and D5-branes.}
\label{table:nonab}
\end{table}
 
\subsection{Engineering a non-Abelian discrete gauge symmetry}
\label{sec:isom-nonab}

The above ingredients allow to engineer a non-Abelian discrete gauge symmetry, in a system with NSNS and RR 3-form fluxes. Consider for simplicity a $\IT^6$ compactification (although the construction may generalize e.g. to torus bundles), parametrized by coordinates $x^4,\ldots, x^9$, and introduce  3-form fluxes ${\ov H}_3\sim p \,dx^6dx^7dx^9$ and ${\ov F}_3\sim p \, dx^7dx^8dx^9$, with the same number of flux quanta for simplicity. As in the previous section, ${\ov H}_3$ breaks the KK $U(1)_{x^9}$ down to $\IZ_p$ (with charged strings given by NS5-branes wrapped on $x^4,x^5,x^8,x^9$, and annihilating on a suitable TN, see first three rows in table \ref{table:nonab}). By S-duality, ${\ov F}_3$ breaks the KK $U(1)_{x^8}$ down to $\IZ_p$ (with charged strings played by D5-branes wrapped on $x^4, x^5, x^6, x^7$, and annihilating on a different TN, see middle three rows in table \ref{table:nonab}). As in section \ref{sec:non-abelian}, the two $\IZ_p$ factors are non-commuting, because crossing the corresponding 4d strings produces a new string, given by a D3-brane wrapped on $x^4$, $x^5$, by the HW effect. The resulting symmetry is a discrete Heisenberg group, with relations $A^p=B^p=1$, $AB=CBA$, and $C^p=1$. The non-Abelianity is analogous to that of section 4.2 in \cite{BerasaluceGonzalez:2012vb} (see also \cite{Gukov:1998kn} for an early 5d realization), with the difference that in the present case the two  $\IZ_p$ factors arise from flux catalysis, rather than from torsion (co)homology. The relation $C^p=1$ follows because the D3-brane 4d strings are associated to a  $\IZ_p$ discrete symmetry induced by the 3-form fluxes (in a way different from section \ref{sec:IIB}, since it requires the presence of non-trivial 1-cycles) through the following Chern-Simons term
\beqa
 \int_{10d} C_4\wedge (F_3\wedge {\ov H}_3 + {\ov F}_3\wedge H_3  )\;\to\; \int_{4d} p \,\hat B_2\,\wedge (\hat F_2 +  \hat F_2')
\eeqa
with $\hat B_2=\int_{x^4 x^5} C_4$, and 4d $U(1)$ field strengths $\hat F_2=\int_{x^8} F_3$ and $\hat F_2'=\int_{x^9}H_3$.

\end{document}